\documentclass[twocolumn]{svjour3}          % twocolumn
\smartqed  % flush right qed marks, e.g. at end of proof
\usepackage{graphicx}
\usepackage{natbib}
\def\nul#1{}
\def \vv#1{\mathbf{#1}}
\def \be {\begin{equation}}
\def \ee {\end{equation}}
\def \k  {s}
\def \K  {k}
\def \p  {0}
\def \s  {1}
\def \a  {2}
\def \b  {1}
\def \ab {2}
\def \bb {{}}
\def \m  {m_{01}}
\def \M  {m_\a}
\def \ij {i}
\def \ji {{(1-\ij)}}
\def \ijk {{i}}
\def \eq {{}}
\def \de {\theta}

\def\ve{\varepsilon}
\def\longitude{argument }
\def\figpath{Figures/}

\def\norm#1{\left\Vert#1\right\Vert}
\def\abs#1{\left|#1\right|}
\def\Frac#1#2{{{\displaystyle\strut#1}\over{\displaystyle\strut#2}}}

\def\crm{\cr\noalign{\medskip}}
\def\trans#1{{}^t{#1}}
\def\moy#1{\left\langle{#1}\right\rangle}
\def\d{{\rm d}}

\def \llabel#1{\label{#1}}
\def\bfx{}
\def\alphaij{\alpha_{2i}}
\def\betaij{\alpha_{1i}}
\def\betaund#1{\alpha_{1{#1}}}

\begin{document}
\title{Tidal evolution of hierarchical and inclined systems}
%\title{Tidal evolution of hierarchical highly-inclined systems}
%\title{Tidal evolution in the secular quadrupolar 3-body problem}
%\title{Tidal evolution of hierarchical planetary systems}
%\subtitle{Application to hierarchical and inclined planetary systems}


\author{A.C.M. Correia         \and
        J. Laskar                 \and
        F. Farago            \and
        G. Bou\'e                
}

%\authorrunning{A.C.M. Correia \and J.Laskar} % if too long for running head

\institute{A.C.M. Correia \at
              Department of Physics, I3N, University of Aveiro,
	      Campus Universit\'ario de Santiago, 3810-193 Aveiro, Portugal \\
             \email{correia@ua.pt}           
}

\date{Received: date / Accepted: date}

\maketitle

\begin{abstract}

We investigate the dynamical evolution of hierarchical three-body systems under
the effect of tides, when the ratio of the orbital semi-major axes is small and
the mutual inclination is relatively large (greater than $20^\circ$).
Using the quadrupolar non-restricted approximation for the gravitational
interactions and the viscous linear model for tides, we derive the averaged equations
of motion in a  vectorial formalism  which is suitable to model
the long-term evolution of a large variety of exoplanetary systems in very
eccentric and inclined orbits.
In particular, it can be used to derive constraints for stellar spin-orbit
misalignment, capture in Cassini states, tidal-Kozai migration, or damping of the mutual inclination.
Because our model is valid for the non-restricted problem, it can be used to
study systems of identical mass or for the outer restricted problem,
such as the evolution of a planet around a binary of stars.
Here, we apply our model to three distinct situations: 1) the HD\,80606 planetary
system, for which we obtain the probability density function distribution for the
misalignment angle, with two pronounced peaks of higher probability around 53$^\circ$
and 109$^\circ$; 2) the HD\,98800 binary system, for which we show that initial prograde orbits
inside the observed disc may become retrograde and vice-versa, only because of tidal
migration within the binary stars; 3) the HD\,11964 planetary system, for which we show
that tidal dissipation combined with gravitational perturbations may lead to a
decrease in the mutual inclination, and a fast circularization of the inner orbit. 

\end{abstract}

%________________________________________________________________

\section{Introduction}

At present, about 50 multi-planet systems have been reported, out of which
roughly 1/3 possess close-in planets with semi-major axis smaller than 0.1\,AU.
%\footnote{http://exoplanet.eu}
%At present, there are 26 %18, 22, 26
%out of 49 multi-planet systems with at least one planet with semi-major axis 
%$ a_\b < 0.3 $\,AU, that is, about 50\%. %0.1, 0.2, 0.3
There are also indications that about half of these stars are likely to have
distant companions \citep{Fischer_etal_2001}. 
{\bfx
In addition, close binary star systems (separation smaller than 0.1\,AU) are
often accompanied by a third star \citep[e.g.][]{DAngelo_etal_2006}.
}
Therefore, although hierarchical systems are considerably different
from our Solar system, they represent a significant fraction of the already known
systems of stars and planets.
These  systems are particularly interesting from a dynamical point of
view, as they can be stable for very eccentric and inclined orbits and thus
present uncommon behaviors.
In particular, they become interesting when the two innermost bodies are
sufficiently close to undergo significant tidal interactions over the age of the
system, since the final outcome of the evolution can be in a configuration
that is totally different from the initial one.

The origin and evolution of the orbital configurations of multi-body systems can 
be analyzed with direct numerical integrations of the full equations of motion, 
but the understanding of the dynamics often benefits from  analytical approximations.
Additionally, tidal effects usually act over very long time-scales and therefore
approximate theories also allow to speed-up the numerical simulations and to
explore the parameter space much more rapidly.
Secular perturbation theories based on series expansions have been used for
hierarchical triple systems.  
{\bfx
For low values of the eccentricity, the expansion of the perturbation 
in series of the eccentricity is very efficient \citep[e.g.][]{Wu_Goldreich_2002},
but this method is not appropriate for orbits that become very eccentric.
An expansion in the  ratio of the semi-major axis $a_\b / a_\a $ is then 
preferred, as in this case exact expressions can be computed for the 
secular system \citep[e.g.][]{Laskar_Boue_2010}. 
}

The development to the second order in  $a_\b / a_\a $, called the quadrupole approximation, was
used by \citet{Lidov_1961,Lidov_1962} and \citet{Kozai_1962} for the restricted inner
problem (the outer orbit is unperturbed).
In this case, the conservation of the normal component of the angular momentum
enables the inner orbit to periodically exchange its eccentricity with
inclination (the so-called Lidov-Kozai mechanism).
There is, however, another limit case to the massive problem, which is the
outer restricted problem (the inner orbit is unperturbed).
\citet{Palacian_etal_2006} have studied this case and discussed the existence
and stability of equilibria in the non-averaged system.
\citet{Farago_Laskar_2010} derived a simple model of the outer restricted case
and described the possible motions of the bodies.
They also looked at the quadrupolar problem of three masses and show how the
inner and outer restricted cases are related to the general case.

{\bfx
For planar problems, the series expansions in  $a_\b / a_\a $ should be conducted to the octopole
order \citep[e.g.][]{Marchal_1990,
Ford_etal_2000,Laskar_Boue_2010}, as the quadrupole approximation  
fails to reproduce the eccentricity oscillations \citep[e.g.][]{Lee_Peale_2003}. 
However, the inclinations of the already known hierarchical systems have not been
yet determined, and it can be assumed that high values for
the eccentricities may also indicate that their mutual inclinations are large as well
\citep[e.g.][]{Laskar_1997,Chatterjee_etal_2008}. 
}

As for Mercury, Venus and the majority of the natural satellites in the Solar
system, close-in bodies undergo significant tidal interactions, resulting that
their spins and orbits are slowly modified. 
The ultimate stage for tidal evolution is the synchronization of the spin and
the circularization of the orbit.
{\bfx
Indeed, the observed mean eccentricity for planets and binary stars with $ a_\b
< 0.1 $\,AU is close to zero within the observational limitations 
\citep[e.g.][]{Pont_etal_2011}.
}
Although tidal effects modify the spin in a much shorter time-scale than they
modify the
orbit, synchronous rotation can only occur when the eccentricity is very close
to zero: the rotation rate tends to be locked with the orbital speed at
the periapsis, because tidal effects are stronger when the two bodies are closer
to each other.
%This is exactly the case observed for Mercury, whose average orbital
%eccentricity is around 0.2, and its rotation is captured in a 3/2 spin-orbit
%resonance \citep{Correia_Laskar_2004}.

During the formation process, the orbital eccentricity can  increase  due to
gravitational scattering, so that the inner bodies become
close enough at periapsis for tidal interactions to occur \citep[e.g.][]{Nagasawa_etal_2008}.
The same gravitational scattering is simultaneously responsible for an increase
of the mutual inclination of the orbits \citep[e.g.][]{Chatterjee_etal_2008}, and
the fact that inclined systems exchange its inclination with the inner's
orbit eccentricity, results that the dissipation in the eccentricity can be
transmitted to the inclination of the orbits, and vice-versa.
The most striking example is that the spin and the orbit can be completely
misaligned \citep[e.g.][]{Pont_etal_2009,Triaud_etal_2010}. 
Previous studies on this subject have been undertaken by
\citet{Eggleton_Kiseleva_2001} for binary stars and by \citet{Wu_Murray_2003}
and \citet{Fabrycky_Tremaine_2007} for a planet in a wide binary.
Despite the success obtained by these works in explaining the observations, they
all used the same set of equations, derived by \citet{Eggleton_Kiseleva_2001},
which is not easy to implement and has been obtained in the frame of the inner
restricted quadrupolar approximation.
As a consequence, their model cannot be applied to a large number of situations,
where the outer orbit cannot be held constant,
such as regular planetary systems or planets around close binaries.
% maybe talk about Mardling & Lin 's work here

In this paper we intend to go deeper into the study of hierarchical three-body
systems, where the innermost bodies undergo tidal interactions.
We do not make any restrictions on the masses of these bodies, and use the
quadrupolar approximation for gravitational interactions with general relativity
corrections.
{\bfx
Our study is then suitable for  binary star systems, planetary
systems, and also for planet-satellite systems.
}
We also consider in our model the full effect on the spins of the two closest
bodies, including the rotational flattening of their figures.
This allows us to correctly describe the precession of the spin axis and
subsequent capture in Cassini states.
We adopt a viscous linear model for tides
\citep{Singer_1968,Mignard_1979}, as it provides simple expressions for the tidal
torques for any eccentricity value.
Since we are interested in the secular behavior, we average the motion
equations over the mean anomalies of the orbits and express them using the
vectorial methods developed by \citet{Boue_Laskar_2006}, \citet{Correia_2009},
and \citet{Tremaine_etal_2009}.

In Section\,\ref{secmodel} we derive the averaged equations of motion that we
use to evolve hierarchical systems by tidal effect. 
In Section\,\ref{secevol} we obtain the secular evolution of the spin and
orbital quantities in terms of reference angles and elliptical elements, that are
useful and more intuitive to understand the outcomes of the numerical simulations.
In Section\,\ref{appexo} we apply our model to three distinct situations of
extra-solar systems: HD\,80606, HD\,98800, and HD\,11964.
Finally, last section is devoted to the conclusions.

\section{The model}

\llabel{secmodel}

We consider here a hierarchical system of bodies composed of a central pair with
masses $m_\p$ and $m_\s$, together with an external companion with mass $\M$.
Both inner bodies are considered oblate ellipsoids with gravity field
coefficients given by $J_{2_\p}$ and $J_{2_\s}$, rotating about the axis of maximal
inertia along the directions $\vv{\hat \k}_\p$ and $\vv{\hat \k}_\s$
(gyroscopic approximation), with rotation
rates $\omega_\p$ and $\omega_\s$, respectively, such that
\citep[e.g.][]{Lambeck_1988}
\begin{equation}
J_{2_\ij} = k_{2_\ij} \frac{\omega_\ij^2 R_\ij^3}{3 G m_\ij} \ ,
\llabel{101220a}
\end{equation}
where $G$ is the gravitational constant, $R_\ij$ is the radius of each body, and
$ k_{2_\ij} $ is the second Love number for potential (pertaining to a perfectly
fluid body).

We use Jacobi canonical coordinates, with $\vv{r_\b} $ being the position of
$m_\s$ relative to $m_\p$, and $ \vv{r_\a} $ the position of $ \M $
relative to the center of mass of $ m_\p $ and $ m_\s $.
We further assume that $|\vv{r_\b}|  \ll  |\vv{r_\a}|$, and 
we shall refer to the orbit of $m_\s$ relative to $m_\p$ as the inner orbit, and
the orbit of $ \M $ relative to the center of mass of $ m_\p $ and $ m_\s $ as
the outer orbit.
In the following, for any vector $\vv{u}$, $\vv{\hat u} = \vv{u} / \norm{\vv{u}}$ is the unit vector. 

\subsection{Conservative motion}
In the 
quadrupolar three-body problem approximation,
the potential energy $U$ of the system is given by 
\citep[e.g.][]{Smart_1953}: 
%\citep{Tisserand_1891,Smart_1953}: 

\begin{eqnarray}
U & =&  
- G \frac{m_\p m_\s}{r_\b} \left( 1 - \sum_{\ij=\p,\s} J_{2_\ij}
 \left(\frac{R_\ij}{r_\b}\right)^2 \!\! P_2 (\vv{\hat r}_\b \cdot \vv{\hat \k}_\ij) 
\right) \crm &&  
- G \frac{\m \M}{r_\a} \left( 1 - \sum_{i=\p,\s} J_{2_\ij} \frac{m_\ij}{\m}
\left(\frac{R_\ij}{r_\a}\right)^2 \!\! P_2 (\vv{\hat r}_\a \cdot \vv{\hat \k}_\ij)
\right) \crm && 
- G \frac{\beta_\b \M}{r_\a} \left(\frac{r_\b}{r_\a}\right)^2 \!\! P_2 (\vv{\hat
r}_\a \cdot \vv{\hat r}_\b) \ , \llabel{090514a}
\end{eqnarray}
where $P_2(x) = (3x^2-1)/2$ is the Legendre polynomial of degree two, and terms
in $(r_\b/r_\a)^3$ and $(R_\ij/r_j)^3$ have been neglected ($\ij,j=\p,\s$). 
We also have $\m = (m_\p + m_\s) $,
$ \beta_\b = m_\p m_\s / \m $,  $ \beta_\a = \m \M / (\m + \M) $, $\mu_\b=G \m$,
and  $\mu_\a=G (\m + \M)$. 

The evolution of the spins can be tracked by the rotational angular
momenta, $ \vv{L}_\ij \simeq C_\ij \omega_\ij \, \vv{\hat \k}_\ij $. 
In turn,
the evolution of the orbits can be tracked by the orbital angular momenta, $\vv{G}_\ij =
\beta_\ij \sqrt{\mu_\ij a_\ij (1-e_\ij^2)} \, \vv{\hat \K}_\ij $  
(where $\vv{\hat \K}_\ij$ is the unit vector $\vv{\hat G}_\ij$), and the
Laplace-Runge-Lenz vector, which points along the major axis 
in the direction of periapsis with  magnitude $e_\b$:
\be
\vv{e}_\b = \Frac{\vv{\dot r}_\b \times \vv{G}_\b}{\beta_\b  \mu_\b} -
\Frac{\vv{r}_\b}{r_\b} \llabel{100119a} \ .
\ee
$a_\ij$ is the semi-major axis (that can also be expressed using the mean
motion, $n_\ij = \sqrt{\mu_\ij / a_\ij^3 }$), 
$e_\ij$ is the
eccentricity, and $C_\ij$ is the principal moment of inertia. The contributions
to the orbits are easily computed from the above potentials as 
\be
\vv{\dot G}_\b = \vv{r}_\b \times \vv{F}_\b \ , \quad
\vv{\dot G}_\a = \vv{r}_\a \times \vv{F}_\a \ , 
\llabel{090514d}
\ee
%\be
%\vv{\dot G}_\b = - \vv{r}_\b \times \vv{\nabla}_{\!\vv{r}_\b} U , \quad
%\vv{\dot G}_\a = - \vv{r}_\a \times \vv{\nabla}_{\!\vv{r}_\a} U , 
%\llabel{090514d}
%\ee
and
\be
\dot \vv{e}_\b = \Frac{1}{\beta_\b \mu_\b} \left( \vv{F}_\b
\times \Frac{\vv{G}_\b}{\beta_\b} + \vv{\dot r}_\b \times \vv{\dot G}_\b \right)
\llabel{100119b}  \ ,
\ee
where $ \vv{F}_\ij = - \vv{\nabla}_{\!\vv{r}_\ij} U' $, 
with $ U' = U + G m_\p m_\s / r_\b + G \m \M / r_\a $.
%\be
%\dot \vv{e}_\b = \Frac{1}{\beta_\b \mu_\b} \left( -\vv{\nabla}_{\!\vv{r}_\b} U'
%\times \Frac{\vv{G}_\b}{\beta_\b} + \vv{\dot r}_\b \times \vv{\dot G}_\b \right)
%\llabel{100119b}  \ ,
%\ee
%where $ U' = U + G m_\p m_\s / r_\b $.

In Jacobi coordinates, the total orbital angular momentum is equal to 
$\vv{G}_\b + \vv{G}_\a $ \citep[e.g.][]{Smart_1953}.
Since the total angular momentum is conserved, the contributions to the spin of
the bodies can be computed from the orbital contributions: 
\be 
\vv{\dot L}_\p + \vv{\dot L}_\s = - (\vv{\dot G}_\b + \vv{\dot G}_\a) \ .
\llabel{090514e}
\ee

Because we are only interested in the secular evolution of the system, we
further average the equations of motion over the mean
anomalies of both orbits (see appendix~\ref{apenA}). 
The resulting equations are
\citep[e.g.][]{Boue_Laskar_2006,Farago_Laskar_2010}:
\begin{eqnarray}
\vv{\dot G}_\b &=& - \gamma (1-e_\b^2) \cos I \, \vv{\hat \K}_\a \times \vv{\hat
\K}_\b + 5 \gamma (\vv{e}_\b \cdot \vv{\hat \K}_\a) \, \vv{\hat \K}_\a \times \vv{e}_\b
\crm &&
- \sum_\ij \betaij \cos \de_\ij \, \vv{\hat \k}_\ij \times \vv{\hat \K}_\b 
 \ , \llabel{090514z2}
\end{eqnarray}
\begin{eqnarray}
\vv{\dot G}_\a &=& - \gamma (1-e_\b^2) \cos I \, \vv{\hat \K}_\b \times \vv{\hat
\K}_\a + 5 \gamma (\vv{e}_\b \cdot \vv{\hat \K}_\a) \, \vv{e}_\b \times \vv{\hat \K}_\a
\crm &&
- \sum_\ij \alphaij \cos \ve_\ij \, \vv{\hat \k}_\ij \times \vv{\hat \K}_\a 
 \ , \llabel{090514z1}
\end{eqnarray}
\begin{eqnarray}
\vv{\dot e}_\b &=& - \Frac{\gamma (1-e_\b^2)}{\norm{\vv{G}_\b}} \left[ \cos I \,
\vv{\hat \K}_\a \times \vv{e}_\b - 2 \, \vv{\hat \K}_\b \times \vv{e}_\b - 5
(\vv{e}_\b \cdot \vv{\hat \K}_\a) \, \vv{\hat \K}_\a \times \vv{\hat \K}_\b \right]
\crm &&
- \sum_\ij \Frac{\betaij}{\norm{\vv{G}_\b}} \left[ \cos \de_\ij \,
\vv{\hat \k}_\ij \times \vv{e}_\b + \frac{1}{2} (1 - 5 \cos^2 \de_\ij) \,
\vv{\hat \K}_\b \times \vv{e}_\b \right] \ , \llabel{090514z3}
\end{eqnarray}
and
\begin{equation}
\vv{\dot L}_\ij = 
-  \betaij \cos \de_\ij \, \vv{\hat \K}_\b \times \vv{\hat \k}_\ij 
- \alphaij \cos \ve_\ij \, \vv{\hat \K}_\a \times \vv{\hat \k}_\ij 
\ , \llabel{090514f}
\end{equation}
where
\be 
\betaij =  \frac{3 G m_\p m_\s J_{2_\ij} R_\ij^2}{2 a_\b^3 (1-e_\b^2)^{3/2}} \ ,
\llabel{090514h}
\ee
\be 
\alphaij = \frac{3 G \M m_\ij J_{2_\ij} R_\ij^2}{2 a_\a^3 (1-e_\a^2)^{3/2}} \ ,
\llabel{090514g}
\ee
\be 
\gamma = \frac{3 G \M \beta_\b a_\b^2}{4 a_\a^3 (1-e_\a^2)^{3/2}} \ ,
\llabel{090514i}
\ee
and
\be
\cos \de_\ij = \vv{\hat \k}_\ij \cdot \vv{\hat \K}_\b \ , \quad
\cos \ve_\ij = \vv{\hat \k}_\ij \cdot \vv{\hat \K}_\a \ , \quad
\cos I   = \vv{\hat \K}_\b \cdot \vv{\hat \K}_\a \ , \llabel{090514j}
\ee
are the direction cosines of the spins and orbits: $ \de_\ij$ is the obliquity
to the orbital plane of the inner orbit, $\ve_\ij$ is the obliquity to the orbital
plane of the outer companion, and $I$ is the inclination between orbital planes. 
They can also be expressed as
\be
\cos \ve_\ij = \cos I \cos \de_\ij + \sin I \sin \de_\ij \cos \varphi_\ij 
\ , \llabel{090521a}
\ee
where $ \varphi_\ij $ is a precession angle. 
%between the projections of $
%\vv{\hat \k}_\ij $ and $ \vv{\hat \K}_\a $ in the plane normal to $ \vv{\hat \K}_\b $. 

\subsection{General relativity correction}

We may add to Newton's equations the dominant contribution from
general relativistic effects. 
{\bfx
These effects are mainly felt by eccentric orbits on close encounters
between the central bodies and contribute to the gravitational force with a small
correction %\citep[e.g.][]{dInverno_1992}
\citep[e.g.][]{Schutz_1985}
\be
\vv{F}_\mathrm{gr} = - \frac{3 \mu_\b \norm{\vv{G}_\b}^2}{\beta_\b c^2 r_\b^4}
\, \vv{\hat r}_\b \ , \llabel{101029a}
\ee
where $ \norm{\vv{G}_\b} = \beta_\b \sqrt{\mu_\b a_\b (1-e_\b^2)} $, and $ c $
is the speed of light. 
% is independent of $\vv{r}$.
}
To this order, the dominant relativistic secular contribution is on the
precession of the periapsis, leaving eccentricity, orientation and semi-major
axis of the orbit unaffected (Eq.\,\ref{100119b}): 
\be
\dot \vv{e}_\b = \frac{3 \mu_\b n_\b}{c^2 a_\b (1-e_\b^2)} \, \vv{\hat \K}_\b
\times \vv{e}_\b \ .
\llabel{101029b}
\ee
%where $n_\ij = \sqrt{\mu_\ij / a_\ij^3 }$ is the mean motion.

\subsection{Tidal effects}

{\bfx
Neglecting the tidal interactions with the external body $m_\a$, the tidal
potential for the inner pair writes \citep[e.g.][]{Kaula_1964}:
}
%\be
%U_T = - G \frac{m_\p m_\s}{r_\b^3} \left( k_{2_\p} \frac{R_\p^5}{r_\a'^3} P_2
%(\vv{\hat r}_\b \cdot \vv{\hat r}_\a') + k_{2_\s} \frac{R_\s^5}{r_\b'^3} P_2
%(\vv{\hat r}_\b \cdot \vv{\hat r}_\b') \right) \ , \llabel{090514b}
%\ee
\be
U_T = - \frac{G}{r_\b^3} \sum_{\ij=\p,\s} k_{2_\ij} m_\ji^2
\frac{R_\ij^5}{r_\b'^3} P_2 (\vv{\hat r}_\b \cdot \vv{\hat r}_\b')  \ ,
\llabel{090514b} 
\ee
where $\vv{r}_\b'$ is the position of the interacting body
at a time delayed of $\Delta t_\ij$. 
{\bfx
The dissipation of the mechanical energy of tides in the body's interior is
responsible for this delay between the initial perturbation and the maximal
deformation.
As the rheology of stars and planets is badly known, the exact dependence of $\Delta
t_\ij$ on the tidal frequency is unknown.
Many different authors have studied the problem and several models have been
developed so far, from the simplest ones to the more complex \citep[for a review
see][]{Correia_etal_2003, Efroimsky_Williams_2009}.
The huge problem in validating one model better than others is the difficulty to
compare the theoretical results with the observations, as the effect of tides
are very small and can only be detected efficiently after long periods of time.
The qualitative conclusions are more or less unaffected, so, for simplicity, we
adopt here a model with constant $\Delta t_\ij$, which can be made linear
%\citep[][]{Singer_1968,Mignard_1979,Surgy_Laskar_1997}:
\citep[][]{Mignard_1979,Surgy_Laskar_1997}:
}
\be
\vv{r}_\b' \simeq \vv{r}_\b + \Delta t_\ij \left(\omega_\ij \vv{\hat \k}_\ij
\times \vv{r}_\b - \vv{\dot r}_\b \right) \ . \llabel{090514c}
\ee

As for the conservative motion, we can obtain the equations of motion
directly from equations (\ref{090514d}), (\ref{100119b}) and (\ref{090514e})
using $U_T$ instead of $U'$ (see
appendix~\ref{apenA}), that is,  
\be
\vv{\dot G}_\a = 0 \ , \quad
\vv{\dot G}_\b = - \vv{\dot L}_\p - \vv{\dot L}_\s \ , \llabel{110503a}
\ee
\begin{eqnarray}
\dot \vv{e}_\b &=& 
%conservative part
\sum_{\ij} \Frac{15}{2} k_{2_\ij} n_\b \left(
\Frac{m_\ji}{m_\ij} \right) \left( \Frac{R_\ij}{a_\b} \right)^5 f_4 (e_\b) 
\, \vv{\hat \K}_\b \times \vv{e}_\b \crm 
%dissipative part
&-& \sum_{\ij} \Frac{K_\ijk}{\beta_\b a_\b^2} 
\left[ f_4(e_\b) \frac{\omega_\ij}{2 n_\b} (\vv{e}_\b \cdot \vv{\hat \k}_\ij) \,
\vv{\hat \K}_\b 
\right. \crm && \left.
- \left( \frac{11}{2} f_4(e_\b) \cos \de_\ij \frac{\omega_\ij}{n_\b} 
- 9 f_5(e_\b) \right) \vv{e}_\b \right]  \ , \llabel{100119h}
\end{eqnarray}
and
\begin{eqnarray}
\vv{\dot L}_\ij &=& K_\ijk \,
n_\b \left[ f_4(e_\b) \sqrt{1-e_\b^2} \frac{\omega_\ij}{2 n_\b} (\vv{\hat
\k}_\ij - \cos \de_\ij \, \vv{\hat \K}_\b) 
\right. \\ && \left.
- f_1(e_\b) \frac{\omega_\ij}{n_\b} \vv{\hat \k}_\ij + f_2(e_\b)
\vv{\hat \K}_\b 
+  \frac{(\vv{e}_\b \cdot \vv{\hat \k}_\ij) (6 + e_\b^2)}{4
(1-e_\b^2)^{9/2}} \frac{\omega_\ij}{n_\b}  \vv{e}_\b
\right] \ , \nonumber \llabel{090514knew}
\end{eqnarray}
where,
\be
K_\ijk = \Delta t_\ij \frac{3 k_{2_\ij} G m_\ji^2 R_\ij^5}{a_\b^6} \ , \llabel{090514m}
\ee
and
\be
f_1(e) = \frac{1 + 3e^2 + 3e^4/8}{(1-e^2)^{9/2}} \ , \llabel{090514n}
\ee
\be
f_2(e) = \frac{1 + 15e^2/2 + 45e^4/8 + 5e^6/16}{(1-e^2)^{6}} \ , \llabel{090514o}
\ee
\be
f_3(e) = \frac{1 + 31e^2/2 + 255e^4/8 + 185e^6/16 + 25e^8/64}{(1-e^2)^{15/2}} \ , 
\llabel{090514p}
\ee
\be
f_4(e) = \frac{1 + 3e^2/2 + e^4/8}{(1-e^2)^5} \ , \llabel{090515d}
\ee
\be
f_5(e) = \frac{1 + 15e^2/4 + 15e^4/8 + 5e^6/64}{(1-e^2)^{13/2}} \ . \llabel{090515e}
\ee

The first term in expression (\ref{100119h}) corresponds to a permanent tidal
deformation, while the second term corresponds to the dissipative contribution.
The precession rate of $ \vv{e}_\b $ about $ \vv{\hat k}_\b $ is usually much
faster than the evolution time-scale for the dissipative tidal effects.
As a consequence, when the eccentricity is constant over a precession
cycle, we can average expression (\ref{090514knew}) over the \longitude 
of the periapsis and get \citep[][appendix~\ref{apenA}]{Correia_2009}:

\begin{eqnarray}
\vv{\dot L}_\ij = - K_\ijk \,
n_\b \left( f_1(e_\b) \Frac{\vv{\hat \k}_\ij + \cos \de_\ij \,
\vv{\hat \K}_\b}{2} \Frac{\omega_\ij}{n_\b} - f_2(e_\b) \vv{\hat \K}_\b 
\right) \ , \llabel{090514k}
\end{eqnarray}

%\section{Main consequences of tidal evolution}
\section{Secular evolution}

\llabel{secevol}

We have presented the equations that rule the tidal evolution of a hierarchical
system of three bodies in terms of a vectorial formalism. 
However, the spin and orbital quantities are better represented by the 
rotation angles and elliptical elements.
The direction cosines (Eq.\ref{090514j}) are obtained from the
angular momenta vectors, since $ \vv{\hat \k}_\ij = \vv{L}_\ij /
\norm{\vv{L}_\ij} $ and $ \vv{\hat \K}_\ij = \vv{G}_\ij / \norm{\vv{G}_\ij} $, as well as the
rotation rate $ \omega_\ij = \vv{L}_\ij \cdot \vv{\hat \k}_\ij / C_\ij $.
The eccentricity and the semi-major axis can be obtained from
$e_\b = \norm{\vv{e}_\b}$ and $a_\b = \norm{\vv{G}_\b}^2 / (\beta_\b^2 \mu_\b (1
- e_\b^2) )$, respectively.

\subsection{Conservative motion}

\subsubsection{Cassini states}

The obliquities are functions of $\vv{G}_\b$, $\vv{G}_\a$, and $\vv{L}_\ij$
(Eq.\,\ref{090514j}).
Their evolution can be obtained from (Eqs.~\ref{090514z2},
\ref{090514z1}, \ref{090514f}) as
\be
\frac{d \cos \de_\ij}{d t} = \frac{\dot
\vv{G}_\b \cdot ( \vv{\hat \k}_\ij - \cos \de_\ij \vv{\hat
\K}_\b)}{\norm{\vv{G}_\b}} + \frac{\dot \vv{L}_\ij \cdot ( \vv{\hat
\K}_\b - \cos \de_\ij \vv{\hat \k}_\ij)}{\norm{\vv{L}_\ij}} \ ,
\llabel{090520b}
\ee
for the obliquity to the orbital plane of the inner orbit, $\de_\ij$. An
identical expression could be obtained for the obliquity to the outer orbit, 
$ \ve_\ij $, replacing $ \vv{G}_\b $ by $ \vv{G}_\a $.
To simplify, we may average the equations over the \longitude of the periapsis
(appendix~\ref{apenA}),
and the norms of $\norm{\vv{G}_\b}$, $\norm{\vv{G}_\a}$, and $\norm{\vv{L}_\ij}$
become constant.
Thus,
\be
\frac{d \cos \de_\ij}{d t} = \frac{d \vv{\hat \K}_\b}{dt} \cdot ( \vv{\hat
\k}_\ij - \cos \de_\ij \vv{\hat \K}_\b) + \frac{d \vv{\hat \k}_\ij}{dt} \cdot (
\vv{\hat \K}_\b - \cos \de_\ij \vv{\hat \k}_\ij) \ , \llabel{090520bbis}
\ee
where
\begin{eqnarray}
\frac{d \vv{\hat \K}_\b}{dt} &=& -\frac{\gamma}{\norm{\vv{G}_\b}}
\left(1+\frac{3}{2}e_\b^2\right)\cos I\, \vv{\hat \K}_\a \times
\vv{\hat \K}_\b 
\\ \nonumber &&
- \sum_i \frac{\betaij}{\norm{\vv{G}_\b}}
\cos \de_\ij\, \vv{\hat \k}_\ij \times \vv{\hat \K}_\b\ ,
\llabel{eq:cass1}
\end{eqnarray}
and
\be
\frac{d \vv{\hat \k}_\ij}{dt} = -\frac{\betaij}{\norm{\vv{L}_\ij}}\cos \de_\ij\,
\vv{\hat \K}_\b \times \vv{\hat \k}_\ij
-\frac{\alphaij}{\norm{\vv{L}_\ij}}\cos \ve_\ij\,
\vv{\hat \K}_\a \times \vv{\hat \k}_\ij\ .
\llabel{eq:cass3}
\ee
{\bfx
This system has {\em a priori} four degrees of freedom associated to the four
precession angles (one for each orbit, and one per solid body). However,
because the total angular momentum is conserved, there 
is only three degrees of freedom. Regular solutions of 
(Eqs.~\ref{090520bbis}, \ref{eq:cass1}, \ref{eq:cass3}) are thus
combination of three eigenmodes. More precisely, these solutions
}
are composed of a uniform rotation of
all the vectors around the total angular momentum, and in the rotating
frame, vectors describe quasi-periodic motions with only 2 proper
frequencies \citep{Boue_Laskar_2006,Boue_Laskar_2009}. 
The bodies are in an equilibrium configuration, usually called
{\em Cassini state}, when the amplitudes of the quasi-periodic motion vanish.
Although analytical approximations of the solutions can be obtained
\citep{Boue_Laskar_2006, Boue_Laskar_2009}, we assume that $\alphaij \ll
\betaij \ll \gamma $, and also $ \norm{\vv{G}_\b} \ll \norm{\vv{G}_\a} $.
Then, equation (\ref{eq:cass1}) simplifies
\be
\frac{d \vv{\hat \K}_\b}{dt} \approx -\frac{\gamma}{\norm{\vv{G}_\b}}
\left(1+\frac{3}{2}e_\b^2\right)\cos I \, \vv{\hat \K}_\a \times
\vv{\hat \K}_\b\ .
\llabel{eq:cass4}
\ee
The evolution of $\vv{G}_\b$ is thus independent of
$\vv{L}_\ij$, and it has a uniform precession motion at frequency $g$ 
around the total orbital angular momentum, %$ \vv{G}_\b + \vv{G}_\a $, 
where
\be
g \approx - \frac{\gamma}{\norm{\vv{G}_\b}} \left(1+\frac{3}{2}e_\b^2\right)
\cos I \ . \llabel{110503c}
\ee
The evolution of the spin-axes is then simpler in the rotating frame
where $\vv{\hat \K}_\b$ and $\vv{\hat \K}_\a$ are constant. 
Since $\betaij \gg \alphaij$, we have for each body
\be
\frac{d \vv{\hat \k}_\ij}{dt} = -\frac{\betaij}{\norm{\vv{L}_\ij}}
\cos \de_\ij\, \vv{\hat \K}_\b \times \vv{\hat \k}_\ij
-g\, \vv{\hat \K}_\ab \times \vv{\hat \k}_\ij \ , \llabel{110503b}
\ee
which leads to
\be
\dot \de_\ij = g \sin I_\bb \sin \varphi_\ij\ ,
\llabel{eq:cass6}
\ee
and
\be
\dot \varphi_\ij = -\left(\frac{\betaij}{\norm{\vv{L}_\ij}} \cos \de_\ij
+ g\cos I_\bb \right)
+ g \sin I_\bb \frac{\cos \varphi_\ij}{\tan \de_\ij} \ .
\llabel{eq:cass7}
\ee
We used the fact that $\vv{\hat \K}_\b$ and $\vv{\hat \K}_\a$ are coplanar. 
Equation (\ref{eq:cass6}) shows that
the obliquity is oscillating around an equilibrium value given by
$\varphi_\ij = 0$ or $\pi$. Stable configurations for the spin can be
found whenever the vectors ($\vv{\hat \k}_\ij$, $\vv{\hat \K}_\b$,
$\vv{\hat \K}_\a$) are coplanar and
precess at the same rate $ g $
\citep[e.g.][]{Colombo_1966,Peale_1969}.
The equilibrium obliquities can be found setting $ \dot
\varphi_\ij = 0 $ (Eq.\,\ref{eq:cass7}), which provides a single relationship
\citep[e.g.][]{Ward_Hamilton_2004}:
\be
\lambda_\ij \cos \de_\ij^\eq  \sin \de_\ij^\eq  + \sin (\de_\ij^\eq  - I_\bb) = 0 \ ,
\llabel{090519b} 
\ee
where $ \lambda_\ij = \betaij / (\norm{\vv{L}_\ij} g) $ is a dimensionless parameter.
The above equation has two or four real roots for $\de_\ij$, which are known by
{\em Cassini states}. 
For nearly coplanar orbits, we have $ I_\bb \sim 0$, and these solutions
are approximately given by: 
\be
%\de_\s^\eq \simeq 
\tan^{-1} \left(\frac{\sin I_\bb}{\cos I_\bb \pm \lambda_\ij} \right) \ , \quad
%\de_\s^\eq \simeq 
\pm \cos^{-1} \left(-\frac{\cos I_\bb}{\lambda_\ij}\right)  \ . \llabel{090519e}
\ee
For a generic value of $I_\bb$, when $ \abs{\lambda_\ij} \ll 1 $ the first expression
gives the only two real roots of equation (\ref{090519b}). On the other hand,
when $\abs{\lambda_\ij} \gg 1 $, we have four real roots 
approximately given by expressions (\ref{090519e}).
%This last situation is the most common for hierarchical systems that we are
%studying here.

\subsubsection{Lidov-Kozai cycles}

\llabel{101223a}

The variations in the mutual inclination between the inner orbit and
the orbit of the external companion can be obtained from the direction cosine
(Eq.\ref{090514j}) in a similar way as the obliquity (Eq.\,\ref{090520b}):
\be
\frac{d \cos I}{d t} = \frac{\dot \vv{G}_\b \cdot ( \vv{\hat \K}_\a - \cos I
\vv{\hat \K}_\b)}{\norm{\vv{G}_\b}} + \frac{\dot \vv{G}_\a \cdot ( \vv{\hat
\K}_\b - \cos I \vv{\hat \K}_\a)}{\norm{\vv{G}_\a}}  \ . \llabel{101221a} 
\ee
Since $ \norm{\vv{G}_\b} \ll \norm{\vv{G}_\a} $, we obtain from expression
(\ref{090514z2}): 
\begin{eqnarray}
\frac{d \cos I}{d t} &=& \frac{5}{2} \frac{\gamma e_\b^2}{\norm{\vv{G}_\b}}
\cos I \sin^2 I \sin 2 \varpi_\b 
\\ \nonumber && 
- \sum_\ij \frac{\betaij \cos
\de_\ij}{\norm{\vv{G}_\b}} (\vv{\hat \k}_\ij \times \vv{\hat \K}_\b ) \cdot
\vv{\hat \K}_\a \ , \llabel{101221b}
\end{eqnarray}
where $ \varpi_\b $ is the \longitude of the periapsis of the inner orbit,
that is, the angle between the line of nodes of the two orbits and the periapsis
of the inner orbit.

On the other hand, the variations in the eccentricity are easily obtained from
the Laplace-Runge-Lenz vector (Eq.\ref{090514z3}):
\be
\dot e_\b = \frac{\vv{\dot e}_\b \cdot \vv{e}_\b}{e_\b} = \frac{5}{2}
\frac{\gamma (1-e_\b^2) e_\b}{\norm{\vv{G}_\b}} 
\sin^2 I \sin 2 \varpi_\b \ . \llabel{101221c}
\ee

Thus, combining expressions (\ref{101221b}) and (\ref{101221c}) and neglecting
the contributions from the rotational flattening (terms in $\betaij$), we
get 
\be
\frac{d \cos I}{d t} = \frac{e_\b \dot e_\b}{(1-e_\b^2)} \cos I
\ , \llabel{101221d'}
\ee
which can be integrated to give
\be
\sqrt{1-e_\b^2} \cos I = h_\b = Cte \ , \llabel{101221d}
\ee
where $h_\b$ is constant in absence of tides.
The above relation shows that an increase in the eccentricity of the inner orbit
must be accompanied by a decrease in the mutual inclination and vice-versa. 
It corresponds to a {\em happy coincidence}, as called by
\citet{Lidov_Ziglin_1976}, which results from the 
fact that in the quadrupolar approximation the potential energy (Eq.\,\ref{090514a})
does not depend on \longitude of the periapsis of the  external body $\varpi_\a$
\citep[e.g.][]{Farago_Laskar_2010}.
As a consequence, the conjugate Delaunay variable $ \norm{\vv{G}_\a} = \beta_\a
\sqrt{\mu_\a a_\a(1-e_\a^2)}$ is  constant, and, as the semi-major axis are
constant in the secular problem, $e_\a$ is also constant.
Indeed, when we consider only the orbital contributions, the total angular
momentum is
\be
\norm{\vv{G}_\b}^2 +\norm{\vv{G}_\a}^2 + 2 \norm{\vv{G}_\b} \norm{\vv{G}_\a}
\cos I = Cte \ . \llabel{eqangmom}
\ee
Since $ \norm{\vv{G}_\a} $ is constant and $ \norm{\vv{G}_\b} \ll
\norm{\vv{G}_\a} $, it remains $ \norm{\vv{G}_\b} \cos I = Cte $, and we
retrieve the result given by equation (\ref{101221d}).
This is no longer true for the octopole or higher order approximations
\citep[e.g.][]{Laskar_Boue_2010}.

The variations in the inclination and eccentricity (Eqs.~\ref{101221b},
\ref{101221c}) become zero when $ \sin 2 \varpi_\b = 0 $, that is, for $ \cos 2
\varpi_\b = \pm 1 $.
When $ \cos 2 \varpi_\b = -1 $ ($\varpi_\b = \pm \pi / 2$) it is possible to
show that $ \dot \varpi = 0 $ has a solution if $ h_\b \le \sqrt{3/5} $
\citep{Lidov_1962,Kozai_1962}.
Thus, when the mutual inclination is greater than $ \arccos
\sqrt{3/5} \simeq 39.23^\circ $ we have a libration regime for the periapsis
about $ \varpi_\b = \pm \pi / 2 $.
In this regime, one can observe large variations in both $ I $ and $ e_\b $,
known by {\em Lidov-Kozai cycles}.
If the inner orbit is initially circular, the maximum eccentricity achieved is
given by $ e_\b = \sqrt{1-(5/3) \cos^2 I} $.

Lidov-Kozai cycles persist as long as the perturbation from the outer body is the
dominant cause of precession in the inner orbit. 
However, additional sources of precession, such as general relativity or
tides, can compensate the libration mechanism and suppress
the large eccentricity/inclination oscillations
\citep[e.g.][]{Migaszewski_Gozdziewski_2009}.
For $ h_\b > \sqrt{3/5} $ the periapsis of the inner orbit is always in a
circularization regime, so there is only small variations in the eccentricity
and inclination.

\subsection{Tidal evolution}

\subsubsection{Spin evolution}

The variation in the body's rotation rate can be computed from
equation (\ref{090514k}) as $\dot \omega_\ij = \vv{\dot L}_\ij \cdot \vv{\hat
\k}_\ij / C_\ij $, giving \citep{Correia_Laskar_2010}:
\be
\dot \omega_\ij = - \frac{K_\ijk \, n_\b}{C_\ij} 
\left( f_1(e_\b) \Frac{1 + \cos^2 \de_\ij}{2} \Frac{\omega_\ij}{n_\b} -
f_2(e_\b) \cos \de_\ij 
% - \frac{e_\b^2 (e_\b^2 + 6) \sin^2 \de_\ij}{8 (1-e_\b^2)^{9/2}} 
%\frac{\omega_\ij}{n_\b} \cos 2 \varpi_\b 
\right) \ . \llabel{090515a}
\ee

For a given obliquity and eccentricity, the equilibrium rotation rate, obtained
when $ \dot \omega_\ij = 0 $, is attained for:
\be
\frac{\omega_\ij^\eq}{n_\b} = \frac{f_2(e_\b)}{f_1(e_\b)} \, \frac{2 \cos
\de_\ij}{1 + \cos^2 \de_\ij} \ , \llabel{090520a}
\ee
Notice, however, that the above expression is only valid for constant eccentricity
(and also constant semi-major axis), or at least if tidal effects modify the
eccentricity faster than other effects.
Indeed, when the eccentricity is forced by the
gravitational perturbations from a companion body, the limit solution of
expression (\ref{090515a}) is no longer given by equation (\ref{090520a}), but
more generally \citep{Correia_Laskar_2004,Correia_Laskar_2009}:
%\be
%\frac{\omega_\ij (t)}{n_\b} = \frac{1}{g_\ij (t)} \left( \frac{\omega_\ij
%(0)}{n_\b} + \frac{K_\ijk}{C_\ij}  \int_0^t f_2 (e_\b(\tau)) \cos
%\de_\ij (\tau) g_\ij (\tau) \, d \tau  \right) \ , \llabel{110104b} 
%\ee
\be
\frac{\omega_\ij (t)}{n_\b} = \frac{K_\ijk}{C_\ij g_\ij (t)}  \int_0^t f_2
(e_\b(\tau)) \cos \de_\ij (\tau) g_\ij (\tau) \, d \tau  \ ,
\llabel{110104b}  
\ee
with
\be
g_\ij (t) = \exp \left( \frac{K_\ijk}{2 C_\ij} \int_0^t f_1 (e_\b(t)) (1 +
\cos^2 \de_\ij (\tau)) \, d \tau
\right) \ . \llabel{110104c} 
\ee
%where we used the approximation $ \cos \de_\ij \approx 1 $ in the above
%expressions (\ref{110104b}) and (\ref{110104c}), since the final obliquity is small. 
It corresponds to a solution that pursues the instantaneous equilibrium rotation
for a given eccentricity (Eq.\ref{090520a}), but delayed and with a smaller
amplitude, depending on the relative strength of tidal effects. 
The stronger these effects are, the shorter is the time delay and closer are the
amplitudes to expression (\ref{090520a}).

In turn, the dissipative obliquity variations are computed by substituting
equation (\ref{090514k}) in (\ref{090520b}) with $ \norm{\vv{L}_\ij} \ll
\norm{\vv{G}_\b}$, giving: 
\be
\dot \de_\ij \simeq \frac{K_\ijk n_\b}{C_\ij \omega_\ij} \sin \de_\ij
\left( f_1(e_\b) \cos \de_\ij \frac{\omega_\ij}{2 n_\b} - f_2(e_\b) \right) 
\ . \llabel{090520d}
\ee

%For a given rotation rate and eccentricity, the equilibrium obliquity, obtained
%when $ \dot \de_\ij = 0 $, is attained for $ \de_\ij = 0 $ if $ \omega_\ij
%/ n_\b \le 2 f_2 (e_\b) / f_1 (e_\b) $, otherwise   
%\be
%\cos \de_\ij = \frac{f_2(e_\b)}{f_1(e_\b)} \frac{2 n_\b}{\omega_\ij} \ .
%\llabel{110104e}  
%\ee

Because of the factor $ n_\b / \omega_\ij $ in the magnitude of the obliquity
variations, for an initial fast rotating body, the 
time-scale for the obliquity evolution is longer than the time-scale
for the rotation rate evolution (Eq.\ref{090515a}). 
As a consequence, it is  expected that the rotation rate approaches its
equilibrium value (Eq.\ref{090520a}) earlier than the obliquity.
Replacing equation (\ref{090520a}) in (\ref{090520d}), we have for
constant eccentricity:
\be
\dot \de_\ij \simeq - \frac{K_\ijk n_\b}{C_\ij \omega_\ij} f_2(e_\b)
\frac{\sin \de_\ij}{1+\cos^2 \de_\ij} \ .
\llabel{090520e}
\ee
We then conclude that, although the initial behavior of the obliquity depends on
the initial rotation rate, tidal effects always end  by decreasing
the obliquity, since $ \dot \de_\ij \le 0 $.
Thus, the final obliquity tends to be captured
in a small obliquity Cassini state (Eq.\ref{090519e}), that is, 
\be
\de_\ij \simeq -\frac{\sin I_\bb}{\lambda_\ij} = \frac{C_\ij \omega_\ij \gamma
}{\betaij\norm{\vv{G}_\b}}
\left(1+\frac{3}{2}e_\b^2\right)
\cos I_\bb \sin I_\bb
\ . \llabel{090519ebis}
\ee

\subsection{Orbital evolution}

The variations in the norm of the orbital angular momentum %$ \norm{\vv{G}_\b} = \beta_\b n_\b a_\b^2 (1-e_\b^2)^{1/2} $, 
can be computed directly from expression (\ref{090514k}),
since  $ \vv{\dot G}_\b = - \vv{\dot L}_\p - \vv{\dot L}_\s $:
\begin{eqnarray}
\frac{d }{d t} \norm{\vv{G}_\b} &=& - \sum_{\ij} \vv{\dot L}_\ij \cdot \vv{\K}_\b
\crm &=& \sum_{\ij} K_\ijk n_\b \left( f_1(e_\b) \cos \de_\ij
\frac{\omega_\ij}{n_\b} - f_2(e_\b) \right) \ . \llabel{101228a} 
\end{eqnarray}
The variations in the eccentricity are easily obtained from the
Laplace-Runge-Lenz vector (Eq.\ref{100119h}):
\begin{eqnarray}
\dot e_\b &=& \frac{\vv{\dot e}_\b \cdot \vv{e}_\b}{e_\b} \crm &=& \sum_{\ij}
\frac{9 K_\ijk}{\beta_\b a_\b^2} \left( \frac{11}{18} f_4(e_\b) \cos \de_\ij 
\frac{\omega_\ij}{n_\b} - f_5(e_\b) \right) e_\b \ ,
\llabel{090515c}
\end{eqnarray}
while the semi-major axis variations are obtained from the eccentricity and the
norm of the orbital angular momentum: 
\begin{eqnarray}
\frac{\dot a_\b}{a_\b} & = & \frac{2 \dot e_\b e_\b}{(1-e_\b^2)} +
 \frac{2 \, \vv{\dot G}_\b \cdot \vv{G}_\b}{\norm{\vv{G}_\b}^2} 
% \crm & = & 
\crm & = & \sum_{\ij} \frac{2 K_\ijk}{\beta_\b a_\b^2} \,
\left( f_2(e_\b) \cos \de_\ij \frac{\omega_\ij}{n_\b} - f_3(e_\b)
\right) \ . \llabel{090515b}
\end{eqnarray}

The inner orbit can either expand or contract, depending on the initial spin of
the bodies. 
Considering for simplicity dissipation in only one component, for  fast initial rotating rates
($ \omega_\ij \gg n_\b $), the semi-major axis and the eccentricity usually
increase, except for retrograde spins ($ \de_\ij > \pi / 2 $). 
The ratio between orbital and spin evolution time-scales is roughly given by 
$ C_\ij / (m_\b a_\b^2) \ll 1 $, meaning that the spin achieves an equilibrium
position much faster than the orbit.
Thus, as the rotation rate decreases, the increasing tendency in the orbital
parameters is reversed when $ d \norm{\vv{G}_\b} /d t = 0 $
(Eq.\,\ref{101228a}),  
\be
\frac{\omega_\ij}{n_\b} \cos \de_\ij = \frac{f_2(e_\b)}{f_1(e_\b)} \ ,
\llabel{110117a}
\ee
that is, when the rotation rate is close to its equilibrium value
(Eq.\,\ref{090520a}).
After an initial increase in the semi-major axis and in the eccentricity,
we can then always expect a contraction of the inner orbit until it becomes
completely circularized. 
The final evolution of the system is then achieved when $ e_\b = 0 $, $
\omega_\ij = n_\b $ (Eq.\ref{090520a}) and $ \de_\ij \simeq -\sin I / \lambda_\ij
$ (Eq.\ref{090519ebis}).

The variations in the mutual inclination can be obtained using expression
(\ref{101221a}) with $ \vv{\dot G}_\b = - \sum_\ij \vv{\dot L}_\ij  $
(Eq.\,\ref{090514k}). 
We have then: 
\be
\frac{d \cos I}{d t} = \sum_{\ij} \frac{K_\ijk \omega_\ij}{2 \norm{\vv{G}_\b}}
f_1 (e_\b) \left( \cos \ve_\ij - \cos I \cos \de_\ij \right)  
\ , \llabel{110104a}  
\ee
or, making use of expression (\ref{090521a}),
\be
\frac{d I}{d t} = - \sum_{\ij} \frac{K_\ijk \omega_\ij}{2 \norm{\vv{G}_\b}} f_1
(e_\b) \sin \de_\ij \cos \varphi_\ij \ . \llabel{101222a}
\ee
Just after the spin of one body is trapped in a small obliquity Cassini state, $ \varphi_\ij =
0 $ and $ \sin \de_\ij $ is constant and given by expression
(\ref{090519ebis}).
Thus, %the mutual inclination can only decrease, since
\be
\frac{d I}{d t} \simeq \sum_{\ij} \frac{K_\ijk \omega_\ij}{2 \lambda_\ij
\norm{\vv{G}_\b}} f_1 (e_\b) \sin I \ . \llabel{101222b}
\ee
%\be
%\frac{d I}{d t} \simeq - \sum_{\ij} \frac{K_\ijk \omega_\ij}{2 \abs{\lambda_\ij}
%\norm{\vv{G}_\b}} f_1 (e_\b) \sin I \ . \llabel{101222b}
%\ee

%Another consequence is that $ \dot \vv{G}_\b = - \dot \vv{L}_\p \simeq 0 $, and
%the quantity $ a_\b (1 - e_\b^2) $ is conserved.
%The final equilibrium semi-major axis is then given by
%\be
%a_e = a_\b (1 - e_\b^2) \ . \llabel{090522c}
%\ee
%However, from this point onwards, the tidal effects on the planet cannot be
%neglected (Eq.\ref{090515b}), and they govern the future evolution of the satellite's
%orbit.
%For $ a_e < a_s $ and $ \de_\p < \pi / 2$, where $ a_s^3 = G m_\p /
%(\omega_\p \cos \de_\p)^2 $, the semi-major axis continues to decrease
%until the planet crashes into the star, while in the remaining situations
%it moves away.

\section{Application to exoplanets}

\llabel{appexo}

In this section we apply the model described in Sect.\,\ref{secmodel} to
three distinct situations of exoplanetary systems:
HD\,80606 (inner restricted problem), HD\,98800 (outer restricted problem), and
HD\,11964 (intermediate non-restricted problem).
{\bfx
We numerically integrate the set of equations (\ref{090514z2}),
(\ref{090514z1}), (\ref{090514z3}), (\ref{090514f}) and (\ref{101029b}) for the
conservative motion, together with equations (\ref{110503a}), (\ref{100119h})
and (\ref{090514knew}) for tidal effects.
}

There exist many systems containing a ``hot Jupiter'' in a wide binary for which one
could apply the present model to illustrate the tidal migration combined with
Lidov-Kozai cycles.
However, we prefer to reproduce the results for HD\,80606 in order to compare our
results (obtained with a non-restricted model) with previous studies by
\citet{Wu_Murray_2003} and \citet{Fabrycky_Tremaine_2007} (obtained with a
restricted model).

The stability of a disc around the multi-binary HD\,98800 system has been
recently studied by \citet{Verrier_Evans_2009} and \citet{Farago_Laskar_2010},
which have isolated two different regimes for the trajectories of the disc.
We then apply our model to an hypothetical planet around the binary stars and
show how transitions between the two regimes are possible, and how
particles in initial prograde orbits may end in retrograde orbits.

We finally use our model to study the HD\,11964 system, which is a hierarchical
planetary system composed of two planets, the inner one in the Neptune-mass
regime and the outer one similar to Jupiter.
There is no determination of the mutual inclination between the two planets, but the
system is stable for very large values, so we analyze its behavior for
different situations.

\subsection{HD\,80606}

In current theories of planetary formation, the region within 0.1~AU of
a protostar is too hot and rarefied for a Jupiter-mass planet to form, so
``hot Jupiters'' likely form further away and then migrate inward.
A significant fraction of ``hot Jupiters'' has been found in systems of binary
stars \citep[e.g.][]{Eggenberger_etal_2004}, suggesting that the stellar companion may play an
important role in the shrinkage of the planetary orbits.
In addition, close binary stars (separation comparable to the stellar
radius) are also often accompanied by a third star.
For instance, \citet{Tokovinin_etal_2006} found that 96\% of a sample of
spectroscopic binaries with periods less than 3~days has a tertiary component.
Indeed, in some circumstances the distant companion induces tidal interactions
in the inner binary by means of the Lidov-Kozai mechanism
(Sect.\,\ref{101223a}), causing the binary semi-major axis to shrink to the
currently observed values \citep[e.g.][]{Eggleton_Kiseleva_2001}. 
The same mechanism has been subsequently proposed to be at the origin of
``hot Jupiters'' as an alternative to migration in a disk
\citep[e.g.][]{Wu_Murray_2003}.

\begin{table}
\caption{Initial parameters for the HD\,80606 system
\citep{Naef_etal_2001,Eggenberger_etal_2004, Wu_Murray_2003}. \llabel{T80606} } 
\begin{center}
\begin{tabular}{|l|c|c|c|c|c|c|} \hline
          & \multicolumn{6}{|c|}{HD\,80606} \\ \hline
parameter & \multicolumn{2}{|c|}{body $\p$} & \multicolumn{2}{|c|}{body $\s$} & \multicolumn{2}{|c|}{body $2$} \\ \hline
$m$ $(M_\odot)$& \multicolumn{2}{|c|}{$ 1.1 $} & \multicolumn{2}{|c|}{$ 0.0037 $} & \multicolumn{2}{|c|}{$ 1.1 $} \\ 
$P_\mathrm{rot}$ $(\mathrm{day})$& \multicolumn{2}{|c|}{$ 20 $} & \multicolumn{2}{|c|}{$ 0.5 $} & \multicolumn{2}{|c|}{$-$}  \\ 
$\de$ $(\mathrm{deg})$ & \multicolumn{2}{|c|}{$ 10 $} & \multicolumn{2}{|c|}{$ 35 $} & \multicolumn{2}{|c|}{$-$}  \\ 
$\varphi$ $(\mathrm{deg})$ & \multicolumn{2}{|c|}{$ 0 $} & \multicolumn{2}{|c|}{$ 0  $} & \multicolumn{2}{|c|}{$-$}  \\ \hline
$R$ $(\times 10^6 \mathrm{m})$ & \multicolumn{2}{|c|}{$ 695 $} & \multicolumn{2}{|c|}{$ 75 $} & \multicolumn{2}{|c|}{$-$}  \\ 
$C / m R^2$ & \multicolumn{2}{|c|}{$ 0.08 $} & \multicolumn{2}{|c|}{$ 0.25 $} & \multicolumn{2}{|c|}{$-$}  \\ 
$k_2$ & \multicolumn{2}{|c|}{$ 0.028 $} & \multicolumn{2}{|c|}{$ 0.51 $} & \multicolumn{2}{|c|}{$-$}  \\ 
$\Delta t$ $(\mathrm{s})$ & \multicolumn{2}{|c|}{$ 0.1 $} & \multicolumn{2}{|c|}{$ 40 $} & \multicolumn{2}{|c|}{$-$}  \\ \hline
parameter & \multicolumn{3}{|c|}{\quad orbit $\b$ \quad \quad} & \multicolumn{3}{|c|}{orbit $\a$} \\ \hline
$a$ $(\mathrm{AU})$& \multicolumn{3}{|c|}{$ 5.0 $} & \multicolumn{3}{|c|}{$ 1000 $}  \\ 
$e$& \multicolumn{3}{|c|}{$ 0.1 $} & \multicolumn{3}{|c|}{$ 0.5 $}  \\ 
$\varpi$ $(\mathrm{deg})$ & \multicolumn{3}{|c|}{$ 0.0 $} & \multicolumn{3}{|c|}{$ - $}  \\ \hline
$I$ $(\mathrm{deg})$&\multicolumn{6}{|c|}{$85.6$}  \\  \hline
%$\Delta \Omega$ $(\mathrm{deg})$&\multicolumn{6}{|c|}{$0.0$}  \\  \hline
\end{tabular}
\end{center}
\end{table}

The HD\,80606 system is composed of two Sun-like stars in a very wide orbit ($
a_\a \sim 1000$\,AU) \citep{Eggenberger_etal_2004}, and a short-period planet in
a very eccentric orbit ($ a_\b = 0.45$\,AU and $e_\b = 0.92 $)
\citep{Naef_etal_2001}.
Because at periapsis the distance to the main star is only 0.036\,AU, the orbit
of the planet is still undergoing tidal evolution, and is thus a perfect example
to test our model. 

In order to compare our results with the previous studies of
\citet{Wu_Murray_2003} and \citet{Fabrycky_Tremaine_2007} we use the same
initial conditions for the planet and the stellar companions:
the planet is initially set in a Jupiter-like orbit with $ a_\b = 5$\,AU, $ e_\b
= 0.1$, and $ I = 85.6^\circ $, while the
stellar companion is supposed to be a Sun-like star at $ a_\a = 1000$\,AU,
and $ e_\a = 0.5 $ (Table\,\ref{T80606}).
In Figure~\ref{kozai} we plot an example of combined tidal-Kozai migration of
the planet HD\,80606\,b.

Prominent eccentricity oscillations are seen from the very beginning and the
energy in the planet's spin is transferred to the orbit increasing the semi-major
axis for the first 0.1\,Gyr (Eq.\,\ref{090515a}).
As the equilibrium rotation is approached around 0.3\,Gyr (Eq.\,\ref{110104b}) the tidal evolution
is essentially controlled by equations (\ref{090515c}) and (\ref{090515b}),
whose contributions are enhanced when the eccentricity reaches high values.
The semi-major axis evolution is executed by apparent
``discontinuous'' transitions precisely because the tidal dissipation is only
efficient during periods of high eccentricity.
As dissipation reduces the semi-major axis, periapsis precession becomes
gradually dominated by general relativity rather than by the third body, and the
periapsis starts circulating as the eccentricity approaches to 0 near 0.5\,Gyr. 
Tidal evolution stops when the orbit is completely circularized.
The final semi-major axis is estimated to about $ a_f = a_\b (1-e_\b^2) \simeq
0.07$~AU, which corresponds to a regular ``hot Jupiter''
\citep{Correia_Laskar_2010B}.

\begin{figure*}
\centering
\includegraphics[width=16cm]{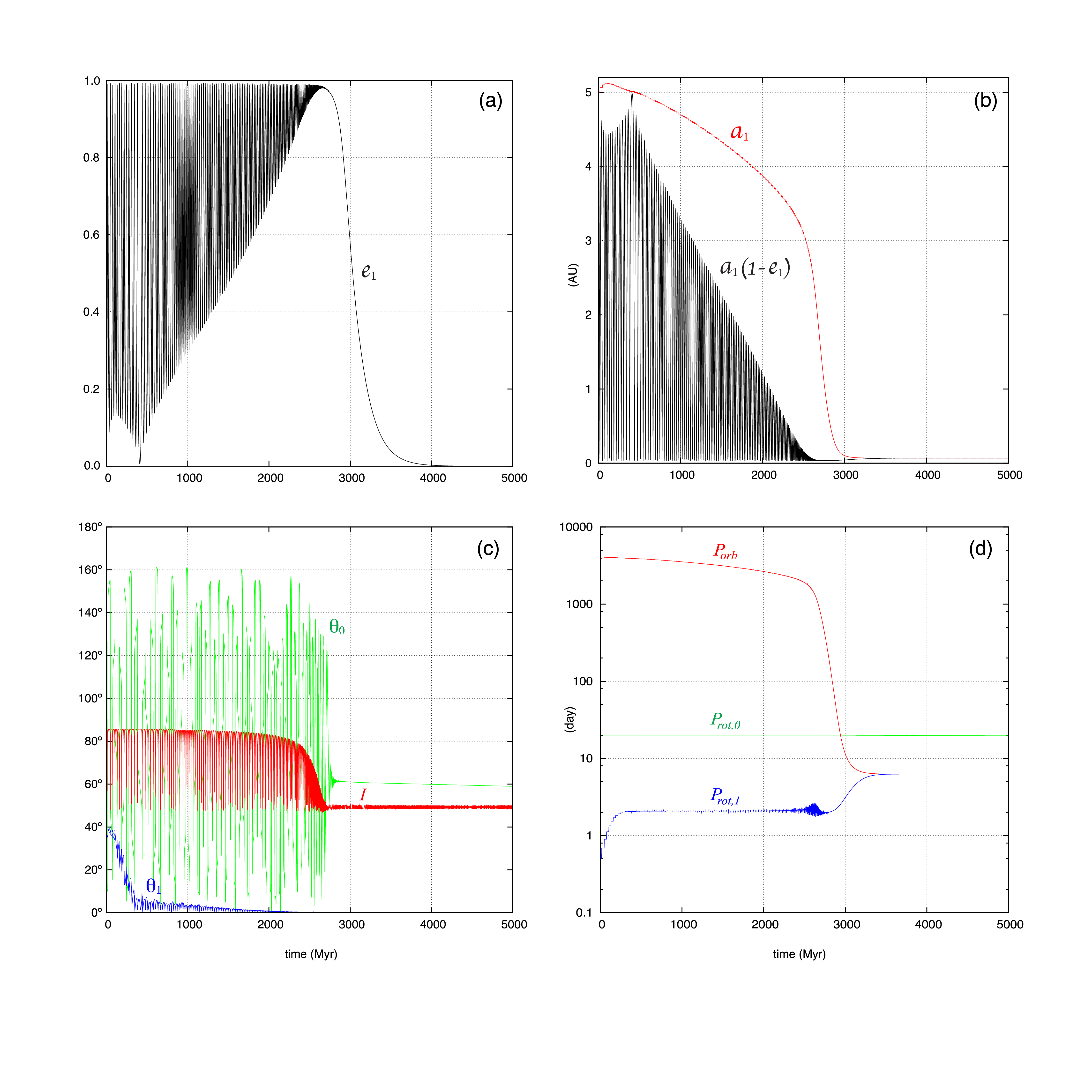}
\caption{Possible evolution of the planet HD\,80606\,b initially in an
  orbit with $ a_\b = 5$\,AU, $ e_\b = 0.1$, and $ I = 85.6^\circ $
  (Table\,\ref{T80606}).
We show the evolution of the eccentricity $e_\b$(a), semi-major axis $a_\b$, and
periapsis $a_\b (1-e_\b)$ (b), mutual inclination $I$, obliquity of the star
$\de_\p$, and obliquity of the planet $\de_\s$ (c), and orbital period
$P_\mathrm{orb}$, rotation period of the star $P_\mathrm{rot,\p}$, and rotation period of the
planet $P_\mathrm{rot,\s}$ (d). Results plot here are identical to those in
\citet{Wu_Murray_2003} and \citet{Fabrycky_Tremaine_2007}, obtained with a
restricted model. 
\llabel{kozai}}
\end{figure*}

The angle between the spin axis of the planet and its orbit (denoted by 
$ \de_\s $) is quickly brought by tides to a small obliquity Cassini state
(Eq.\ref{090519ebis}).
On the other hand, the angle between the spin axis of the star and the orbit of the
planet (denoted by $\de_\p$) is not tidally evolved, since the dissipation
within the star is much smaller than the dissipation within the planet 
($ k_{2_\p} \Delta t_\p \ll k_{2_\s} \Delta t_\s $, Table\,\ref{T80606}). 
As a consequence, capture of the spin of the star in a Cassini state does not
occur during the time length of the simulations, but one can observe large
oscillations corresponding to the variations in the orientation of the orbital
plane of the planet.
Indeed, as long as $ \betaund\p \ll \gamma $, the angle
between the spin of the star and the outer companion (denoted by $\ve_\p$) is
approximately constant (Eq.\ref{110503b}), and thus $|\ve_\p - I | \le \de_\p
\le \ve_\p + I $ (Eq.\,\ref{090521a}).
In our simulation $\ve_\p = 85.6^\circ - 10^\circ = 75.6^\circ$
(Table\,\ref{T80606}), so we approximately observe that $10^\circ \le \de_\p
\le 161^\circ $ (Figure\,\ref{kozai}c).

As the semi-major axis decreases, the precession of the planet's orbit becomes
progressively dominated by the equatorial bulge of the star, so that $ \betaund\p
\sim \gamma $ around 2.8\,Gyr (Eq.\,\ref{eq:cass1}).
From that point, the orbit of the planet precesses around the spin of the star
and $\de_\p$ becomes constant, retaining a value between $10^\circ$ and
$160^\circ$.
One then expects to observe a misalignment between the spin of the star and
the orbit of the planet.
{\bfx
In the simulation shown in Figure\,\ref{kozai}, we obtain $\de_\p \approx
60^\circ$, but this value depends on the initial configuration of the system.

In order to get a statistical distribution of the final misalignment we have
integrated a series of 40\,000 systems with the
same initial conditions from Table\,\ref{T80606} except for the obliquity of the
star $ \de_\p = 0^\circ $, and for the mutual
inclination $I$, which ranged from $\pm 84.3^\circ$ to $90^\circ$ (on an evenly
spaced grid of $ - 0.1 \le \cos I \le +0.1 $). %, as in \citet{Fabrycky_Tremaine_2007}.
A histogram with the results of these simulations  is given in Figure\,\ref{histogram}, with a bin
width of  $1^\circ$.
Our results do not differ much from the histogram obtained by
\citet{Fabrycky_Tremaine_2007}, except that we observe two pronounced peaks of
higher probability around $\de_\p \approx 53^\circ$ and $\de_\p \approx
109^\circ$.
They performed 1\,000 simulations distributed in bins widths of $10^\circ$,
which partially explains this difference, but the main reason is
the fact that they adopted for the rotation of the star $ P_{\mathrm{rot},\p} =
10 $\,h, while we used a value closer to the Sun's rotation period.
Because we executed so much simulations our histogram is close to the
probability density function distribution for the misalignment.
}
Observations of the Rossiter-McLaughlin anomaly for the HD\,80606 star
reinforced the hypothesis of spin-orbit misalignment in this system (alignment
excluded at $ > 95\%$ level), with a positive median projected angle of
$50^\circ$ \citep{Pont_etal_2009}.
These observations are in perfect agreement with our simulations.

\begin{figure}
\centering
\includegraphics[width=8cm]{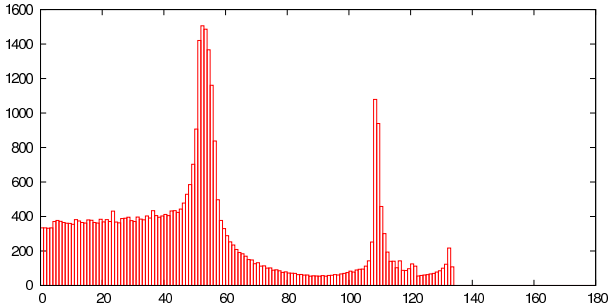}
\caption{Histogram of the final distribution of the misalignment angle
$\de_\p$. We have integrated a series of 40\,000 systems with the
same initial conditions from Table\,\ref{T80606} except for the obliquity $
\de_\p = 0^\circ $, and 
inclination $I$, which range from $\pm 84.3^\circ$ to $90^\circ$. % (an evenly spaced grid of $ - 0.1 \le \cos I \le +0.1 $).
We observe two pronounced peaks of higher probability around $\de_\p \approx
53^\circ$ and $\de_\p \approx 109^\circ$, which is consistent with the
observations of the Rossiter-McLaughlin anomaly for the HD\,80606 system 
\citep{Pont_etal_2009}.
\llabel{histogram}}
\end{figure}

Although we used the full quadrupolar problem, while previous studies
\citep{Wu_Murray_2003,Fabrycky_Tremaine_2007} used the inner restricted problem
(where the outer orbit is considered constant), we retrieve similar results and
then verify that their approximation is appropriate for this specific situation.

\subsection{HD\,98800}

\begin{table}
\caption{Initial parameters for the HD\,98800\,B system \citep{Torres_etal_1995,Tokovinin_1999}. \llabel{T98800} } 
\begin{center}
\begin{tabular}{|l|c|c|c|c|c|c|} \hline
          & \multicolumn{6}{|c|}{HD\,98800\,B} \\ \hline
parameter & \multicolumn{2}{|c|}{body $\p$} & \multicolumn{2}{|c|}{body $\s$} & \multicolumn{2}{|c|}{body $2$} \\ \hline
$m$ $(M_\odot)$& \multicolumn{2}{|c|}{$ 0.699 $} & \multicolumn{2}{|c|}{$ 0.582 $} & \multicolumn{2}{|c|}{$ 0.001 $} \\ 
$P_\mathrm{rot}$ $(\mathrm{day})$& \multicolumn{2}{|c|}{$ 20 $} & \multicolumn{2}{|c|}{$ 0.5 $} & \multicolumn{2}{|c|}{$-$}  \\ 
$\de$ $(\mathrm{deg})$ & \multicolumn{2}{|c|}{$ 10 $} & \multicolumn{2}{|c|}{$ 35 $} & \multicolumn{2}{|c|}{$-$}  \\ 
$\varphi$ $(\mathrm{deg})$ & \multicolumn{2}{|c|}{$ 0 $} & \multicolumn{2}{|c|}{$ 0  $} & \multicolumn{2}{|c|}{$-$}  \\ \hline
$R$ $(\times 10^6 \mathrm{m})$ & \multicolumn{2}{|c|}{$ 758 $} & \multicolumn{2}{|c|}{$ 591 $} & \multicolumn{2}{|c|}{$-$}  \\ 
$C / m R^2$ & \multicolumn{2}{|c|}{$ 0.08 $} & \multicolumn{2}{|c|}{$ 0.08 $} & \multicolumn{2}{|c|}{$-$}  \\ 
$k_2$ & \multicolumn{2}{|c|}{$ 0.028 $} & \multicolumn{2}{|c|}{$ 0.028 $} & \multicolumn{2}{|c|}{$-$}  \\ 
$\Delta t$ $(\mathrm{s})$ & \multicolumn{2}{|c|}{$ 100 $} & \multicolumn{2}{|c|}{$ 100 $} & \multicolumn{2}{|c|}{$-$}  \\ \hline
parameter & \multicolumn{3}{|c|}{\quad orbit $\b$ \quad \quad} & \multicolumn{3}{|c|}{orbit $\a$} \\ \hline
$a$ $(\mathrm{AU})$& \multicolumn{3}{|c|}{$ 0.983 $} & \multicolumn{3}{|c|}{$ 5.2 $}  \\ 
$e$& \multicolumn{3}{|c|}{$ 0.785 $} & \multicolumn{3}{|c|}{$ 0.5 $}  \\ 
$\varpi$ $(\mathrm{deg})$ & \multicolumn{3}{|c|}{$ 82.0 $} & \multicolumn{3}{|c|}{$ - $}  \\ \hline
$I$ $(\mathrm{deg})$&\multicolumn{6}{|c|}{$20.0$}  \\   \hline
%$\Delta \Omega$ $(\mathrm{deg})$&\multicolumn{6}{|c|}{$82.0$}  \\  \hline
\end{tabular}
\end{center}
\end{table}

\begin{figure*}
\centering
\includegraphics[width=16cm]{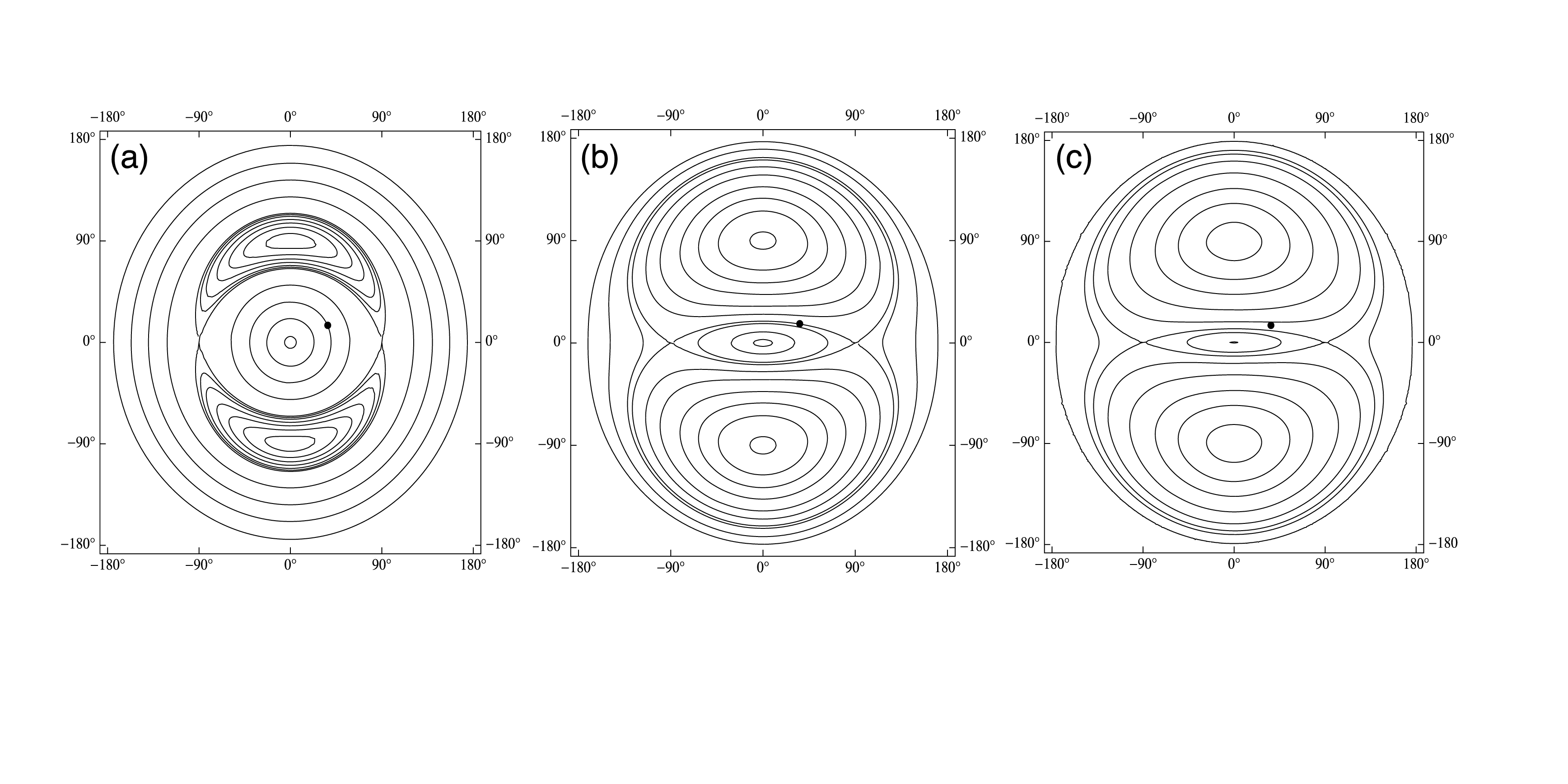}
\caption{Energy levels in the ($ I \sin \varpi $, $ I \cos \varpi $) plane for
different values of the eccentricity of the B~binary HD\,98800, $ e_\s = 0.5 $ (a), $ e_\s = 0.785 $
(b), and $ e_\s = 0.9 $ (c) \citep{Farago_Laskar_2010}. The dot marks the initial
position of the outer orbit (Table\,\ref{T98800}). \llabel{farago}}
\end{figure*}

HD\,98800 is an interesting and unusual system of four 10\,Myr old post T-Tauri~K
stars: two spectroscopic binaries A and B in orbit about one another
\citep{Torres_etal_1995, Kastner_etal_1997}. 
Moreover, it has a large infrared excess attributed to a circumbinary disc around the B
pair \citep{Koerner_etal_2000,Furlan_etal_2007}.
Substantial extinction towards this pair suggests that it is observed
through some of this material \citep{Tokovinin_1999}.
An apparent absence of CO molecular gas in the disc combined with infrared spectrum
modeling indicate that this is a T-Tauri transition disc that is just reaching
the debris disc stage, with a collisional cascade having been recently initiated
\citep[e.g.][]{Furlan_etal_2007}. 
The orbits of the stars are all highly eccentric and inclined, creating a dynamical
environment unlike almost all other known debris discs \citep{Tokovinin_1999}.  
The dust disc is generally agreed to be an annulus around the B~binary, but the
exact structure varies from model to model. 
\citet{Koerner_etal_2000} estimate a coplanar narrow ring outwards of about
$\sim 5$\,AU from the two stars.
However, \citet{Boden_etal_2005} argue that the line of sight extinction means
that the disc cannot be coplanar unless it is very flared.

\citet{Verrier_Evans_2009} investigated numerically the stability of
a family of particles at large inclinations around the B~binary, which remain
stable even under the perturbation of an outer third stellar companion. 
The results show that the Lidov-Kozai mechanism of the outer star is disrupted by a
nodal libration induced by the inner binary pair on a shorter time-scale.
An analytical study by \citet{Farago_Laskar_2010} confirmed that equilibria at
large inclination circumbinary orbits exist for the outer restricted quadrupolar
problem.
This raises the possibility that planets and asteroids with large inclination
may survive in multi-stellar systems.

In this section we test the evolution of a Jupiter-like mass planet
around the B~binary in an orbit that is presently occupied by the dust disc.
Since we average the orbit of the planet over the mean longitude, the orbital
evolution of this planet can also be regarded as the orbital evolution of
the particles in the disc.
The HD\,98800 system is still very young, and the B~binary stars are very close
to each other, so one can expect that the two components
will undergo significant tidal evolution throughout its life.
In order to speed-up the simulations we assume that both stars
experience intense tidal dissipation, with $\Delta t = 10^2$\,s
(Table\,\ref{T98800}). 
%In order to speed-up the simulations we assume that one of the stars
%experiences intense tidal dissipation, with a ratio $\sim 10^3$ between the
%time lags of the two stars (Table\,\ref{T98800}).
Because the gravitational effects of the A~binary do not destabilize the system,
we do not consider its presence in the simulations.

The inclination of the B~binary with respect to the line of sight is estimated
to be about $23^\circ$ \citep{Boden_etal_2005}.
The inclination of the disc with respect to the B~binary is unknown, but one can
contest its coplanarity, since substantial extinction of the light implies that
the system is observed through some of this material \citep{Tokovinin_1999}.
Therefore, we assume that the inclination is about $20^\circ$
(Table\,\ref{T98800}).
The \longitude of the periapsis and the longitude of the node were also
determined with respect to the plane of the sky \citep{Boden_etal_2005}, but
again, we miss the value of these two quantities measured with respect to the
plane of the disc. 
As a consequence, $\varpi_\s$ is a free parameter for the
planet/disc in our simulation.
The choice of this parameter is critical, as it places the initial
system in a different regime of libration, and
it also determines the libration amplitude (Fig.\,\ref{farago}b).
%As a consequence, $\varpi_\s$ and $\Delta \Omega$ are free parameters for the
%planet/disc in our simulation.
%The choice of these two parameters is critical, as it places the initial
%system in a different regime of libration, and
%it also determines the libration amplitude (Fig.\,\ref{farago}b).
Indeed, we can differentiate two kinds of regimes \citep{Farago_Laskar_2010}: 
closed trajectories where the orbital angular moment of the planet $\vv{\K}_\a$
precesses around the orbital angular momentum of the binary $\vv{\K}_\b$ (or its
opposite $-\vv{\K}_\b$); or closed trajectories where the orbital angular momentum
of the planet precesses around the direction of the periapsis of the binary
$\vv{e}_\b $ (or its opposite $- \vv{e}_\b $).
In the first situation the inclination is strictly inferior to $90^\circ$ 
(or strictly superior to $90^\circ$, corresponding to a retrograde orbit),
while in the second case the inclination oscillates around $\pm 90^\circ$.

In our simulation, we use $\varpi_\s = 82^\circ$, as this value places the
initial orbit of the planet at the edge between the two different regimes of
behavior (Fig.\,\ref{farago}b). 
%In our simulation, we use $\varpi_\s = 0^\circ$ and $\Delta \Omega =
%82^\circ$, as these values place the inital orbit of the planet at the edge
%between the two different regimes of behavior (Fig.\,\ref{farago}b).
Initially, $\vv{\K}_\a$ is precessing around $\vv{\K}_\b$ and the mutual
inclination oscillates between $20^\circ$ and nearly $90^\circ$.
As the system evolves, the norm of the orbital angular momentum of the planet
remains constant, but the norm of the orbital angular momentum of the binary is modified by
exchanges with the rotational angular momentum of the stars.
Since we have considered $\omega_\s \gg \omega_\p $ (Table\,\ref{T98800}) in
our example the initial exchanges are mainly between the binary's orbit and the spin of
the star~$\s$. 
%Since we have considered $\Delta t_\s \gg \Delta t_\p $ (Table\,\ref{T98800}) in
%our example the exchanges are only between the binary's orbit and the spin of
%the star~$\s$. 

During the first stages of the evolution $\omega_\s \gg n_\b $ and
the norm of the orbital angular momentum of the binary increases
(Eq.\,\ref{101228a}), as well as the eccentricity and the semi-major axis
(Eqs.~\ref{090515c}, \ref{090515b}). 
As a consequence, the separatrix between the two regimes 
contracts (Fig.\,\ref{farago}c).
Since the norm of the orbital angular momentum of the planet remains constant
(the planet is too far to undergo dissipation), the orbit of the planet crosses the
separatrix and switches to the regime of precession around the direction of the
periapsis of the binary $\vv{e}_\b $. 
In our simulation this transition occurs shortly after 80~Myr since we placed
the initial system very close to the separatrix (Fig.\,\ref{F98800if}a).

\begin{figure*}
\centering
\includegraphics[width=16cm]{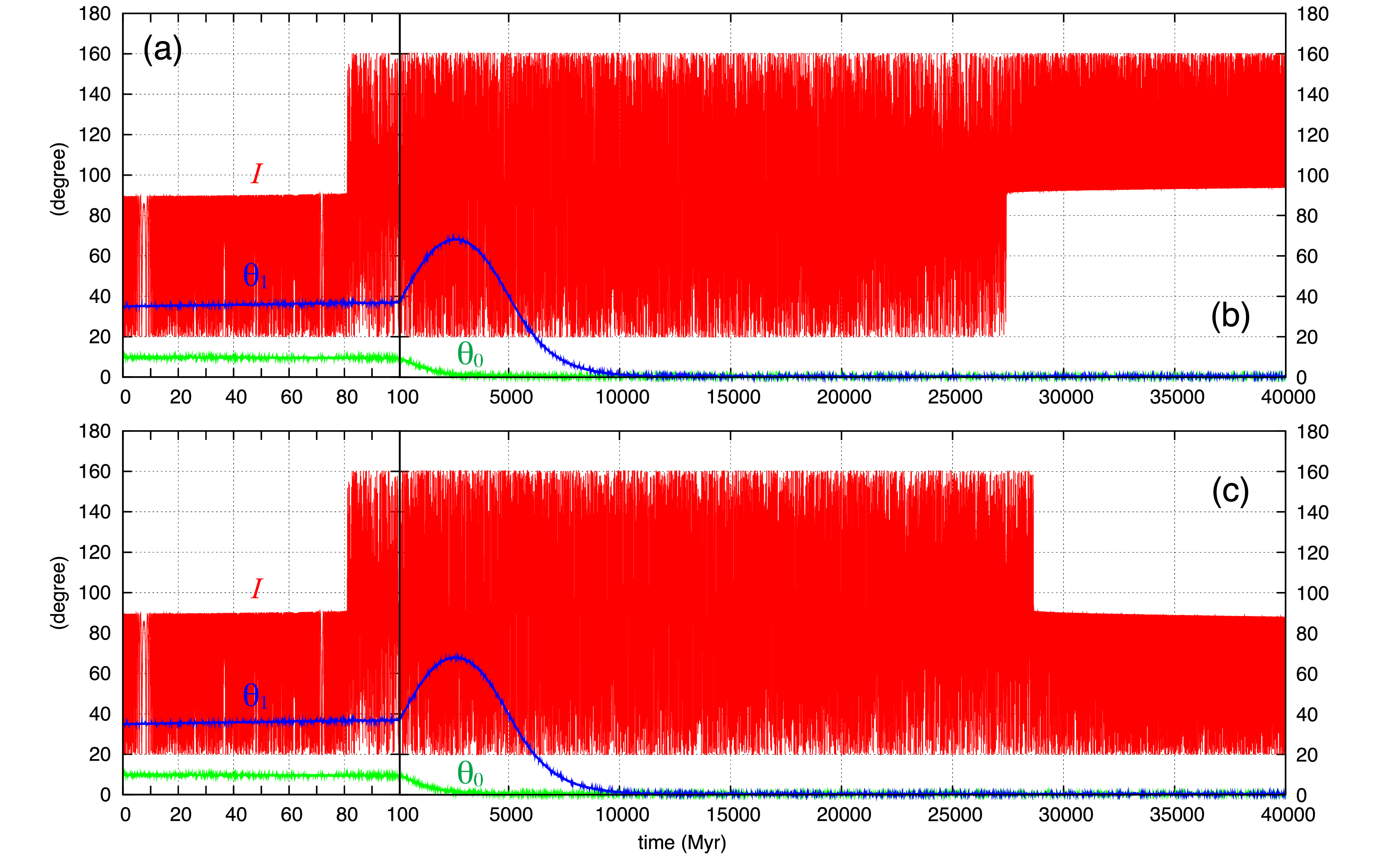}
\caption{Possible evolutions for a $\M \approx 10^{-3} M_\odot$ planet or disc
at $ a_\a = 5.2 $\,AU and $e_\a=0.5$, initially in a prograde orbit around the
binary HD\,98800\,B (Table\,\ref{T98800}). During the first evolutionary stages
the orbit of the binary expands and the planet changes of regime after about
80\,Myr (a). As the binary orbit shrinks, the planet changes of regime again
(somewhere between 25 and 30\,Gyr), but depending when the planet crosses the
separatrix between regimes, its orbit may become retrograde (b) or prograde
(c). \llabel{F98800if}}
\end{figure*}

According to expression (\ref{090520d}) the obliquity of the star $\de_\s$
also increases, but the rotation rate decreases (Eq.\,\ref{090515a}). 
%until an equilibrium is achieved close to the value given by expression (\ref{090520a}).
Around 2\,Gyr, the obliquity begins to decrease and the orbit of the binary
initiates its contraction.
In general, this contraction is much slower than the previous expansion because
the spin is already near an equilibrium position and the major source of
dissipation becomes the orbital energy.
As the eccentricity is  damped, the separatrix between the two regimes expands
and  catches again the planet's orbit (Fig.\,\ref{farago}b).
After the second separatrix transition, the orbital angular momentum of
the planet precesses again around the orbital angular momentum of the binary or
its opposite $\pm \vv{\K}_\b$, depending on the $\varpi_\b$ value at the
time the transition occurs.
%its opposite $\pm \vv{\K}_\b$, depending on the $\Delta \Omega$ value at the
%moment the transition occurs.
We have performed two different simulations, with a $10^{-2}$ difference in the
initial obliquity of the star $\de_\p$, and each one led to the a different final
evolution scenario (Fig.\,\ref{F98800if}b,c).

Subsequent tidal evolution of the binary orbit tends to circularize it.
When the eccentricity reaches zero, the oscillations in the mutual inclination
are also damped and the two angular momenta precess at constant inclination
and speed (Fig.\,\ref{farago}a).
%The final inclination in our simulation would be about $20^\circ ??$ or
%$160^\circ ??$ for retrograde orbits.
Tidal evolution in the inner binary is thus a very efficient mechanism of
transforming initial prograde (or retrograde) orbits in retrograde (or prograde)
orbits, although the norm of the angular momentum of the planet does not change.

\subsection{HD\,11964}

The planetary system around HD\,11964 is composed of two planets with minimum
masses $m_\s = 25\,M_\oplus$ (planet $c$) and $\M = 0.62\,M_{\rm Jup}$ (planet
$b$), at $a_\b = 0.229 $\,AU and $a_\a = 3.16 $\,AU, respectively
\citep{Butler_etal_2006,Wright_etal_2009}. 
%It is member of a wider binary with a separation $\sim 1500 $\,AU
%\citep{Allen_etal_2000}, but we do not consider here the presence of the
%stellar companion, although it may influence the secular dynamics of the system.
\citet{Veras_Ford_2010} performed extensive n-body simulations for this system
and concluded that it is always stable for mutual inclinations $ I < 60^\circ $
(or $ I > 120^\circ$), stable up to 85\% for $ I < 75^\circ $ (or $ I >
105^\circ $), and stable up to 25\% for $ 75^\circ < I < 105^\circ $.
It is a system for which stability is possible for a wide range of
mutual inclinations and thus perfect to apply our secular model.

Since the ratio between the semi-major axis of the two planets is 
$a_\a / a_\b \sim 14$, 
the quadrupolar approximation is well suited, although octopole order terms in
the development of the potential energy (Eq.\,\ref{090514a}) may cause 
variations and exchanges between the eccentricities of both planets. 
In order to test the quality of our secular model in the case of hierarchical
planetary systems of this kind, we performed some direct n-body numerical
simulations for the HD\,11964 system and compared with the quadrupolar secular model. 
Results for the initial parameters in Table~\ref{T11964} are shown in
Figure~\ref{nbodysims}.
We observe slightly additional variations in the amplitudes and precession rates
of the eccentricity and inclination, but the general long-term behavior of the 
system remains essentially the same.

\begin{table}
\caption{Initial parameters for the HD\,11964 system
\citep{Butler_etal_2006,Wright_etal_2009}. \llabel{T11964} } 
\begin{center}
\begin{tabular}{|l|c|c|c|c|c|c|} \hline
          & \multicolumn{6}{|c|}{HD\,11964} \\ \hline
          & \multicolumn{2}{|c|}{body $\p$} & \multicolumn{2}{|c|}{body $\s$} & \multicolumn{2}{|c|}{body $2$} \\ 
parameter & \multicolumn{2}{|c|}{($star$)} & \multicolumn{2}{|c|}{$(c)$} & \multicolumn{2}{|c|}{$(b)$} \\ \hline
$m$ $(M_{Jup})$& \multicolumn{2}{|c|}{$ 1178 $} & \multicolumn{2}{|c|}{$ 0.0788 $} & \multicolumn{2}{|c|}{$ 0.622 $} \\ 
$P_\mathrm{rot}$ $(\mathrm{day})$& \multicolumn{2}{|c|}{$ 20 $} & \multicolumn{2}{|c|}{$ 0.5 $} & \multicolumn{2}{|c|}{$-$}  \\ 
$\de$ $(\mathrm{deg})$ & \multicolumn{2}{|c|}{$ 10 $} & \multicolumn{2}{|c|}{$ 35 $} & \multicolumn{2}{|c|}{$-$}  \\ 
$\varphi$ $(\mathrm{deg})$ & \multicolumn{2}{|c|}{$ 0 $} & \multicolumn{2}{|c|}{$ 0  $} & \multicolumn{2}{|c|}{$-$}  \\ \hline
$R$ $(\times 10^6 \mathrm{m})$ & \multicolumn{2}{|c|}{$ 695 $} & \multicolumn{2}{|c|}{$ 75 $} & \multicolumn{2}{|c|}{$-$}  \\ 
$C / m R^2$ & \multicolumn{2}{|c|}{$ 0.08 $} & \multicolumn{2}{|c|}{$ 0.25 $} & \multicolumn{2}{|c|}{$-$}  \\ 
$k_2$ & \multicolumn{2}{|c|}{$ 0.028 $} & \multicolumn{2}{|c|}{$ 0.50 $} & \multicolumn{2}{|c|}{$-$}  \\ 
$\Delta t$ $(\mathrm{s})$ & \multicolumn{2}{|c|}{$ 0.1 $} & \multicolumn{2}{|c|}{$ 100 $} & \multicolumn{2}{|c|}{$-$}  \\ \hline
parameter & \multicolumn{3}{|c|}{$\,$ orbit $\b$ ($c$) $\,$} & \multicolumn{3}{|c|}{orbit $\a$ ($b$)} \\ \hline
$a$ $(\mathrm{AU})$& \multicolumn{3}{|c|}{$ 0.229 $} & \multicolumn{3}{|c|}{$ 3.16 $}  \\ 
$e$& \multicolumn{3}{|c|}{$ 0.300 $} & \multicolumn{3}{|c|}{$ 0.041 $}  \\ 
$\varpi$ $(\mathrm{deg})$ & \multicolumn{3}{|c|}{$ 0.0 $} & \multicolumn{3}{|c|}{$ - $}  \\ \hline
$I$ $(\mathrm{deg})$&\multicolumn{6}{|c|}{$30.0$}  \\  \hline
\end{tabular}
\end{center}
\end{table}

Though the present age of the star is estimated to be about 10\,Gyr
\citep{Saffe_etal_2005}, we start the system as if it has been recently
formed. We assume the present orbits as the original ones, and an
initial rotation period of the inner planet of about 10\,h (Table\,\ref{T11964}).
Because tidal effects on the inner planet are strong, the choice of
the initial spin is irrelevant, as it evolves to an equilibrium
configuration in about one million years (Figure\,\ref{F11964if}a).
Nevertheless, this procedure allows us to understand simultaneously the initial
behavior of the system and also its future evolution.
In addition, we arbitrarily assume a mutual inclination $ I = 30^\circ $.
This inclination value sets the system well outside a coplanar configuration,
but yet below the critical value of about $39^\circ$ that allows Lidov-Kozai cycles.

\begin{figure}
\centering
\includegraphics[width=8cm]{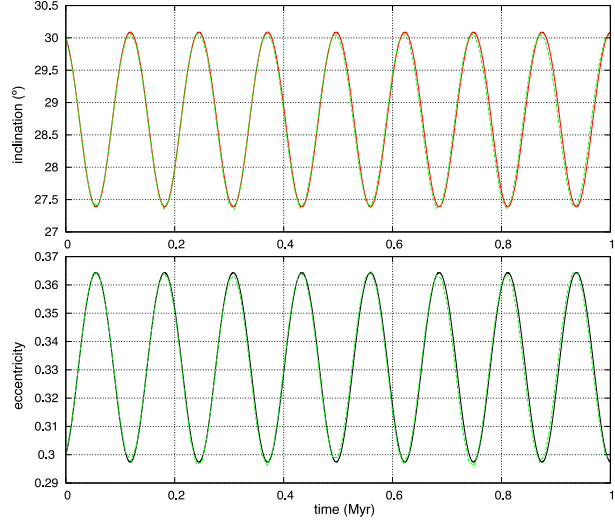}
\caption{Evolution of the HD\,11964\,c inclination (top) and eccentricity
(bottom) during 1\,Myr, starting with the orbital solution from
\citet{Wright_etal_2009} and $I = 30^\circ$ (Table\,\ref{T11964}).
The solid lines curves are the values obtained with the quadrupolar secular
model, while the dashed green lines are the complete solutions obtained with a
n-body numerical model. \llabel{nbodysims}}
\end{figure}

\begin{figure*}
\centering
\includegraphics[width=16cm]{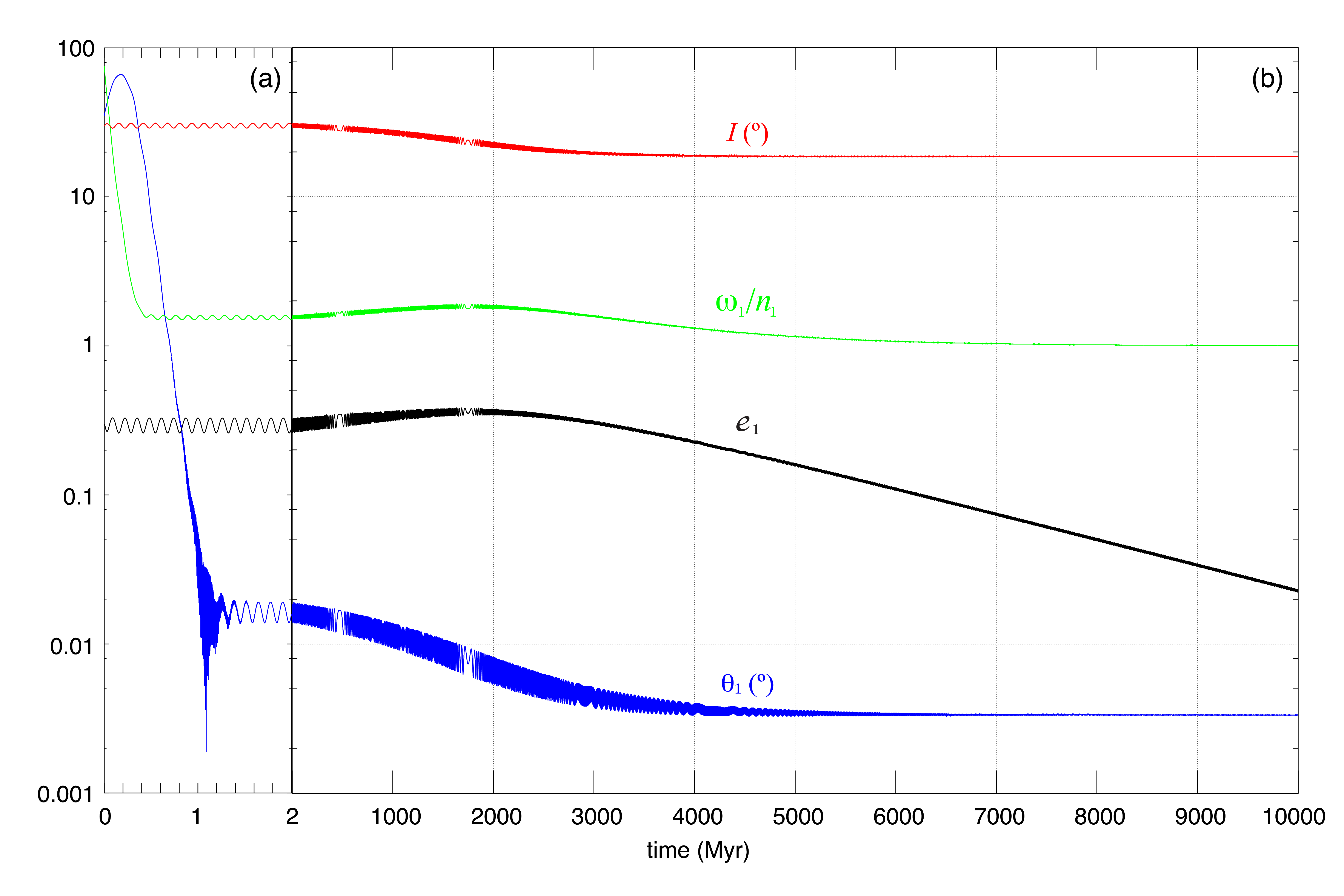}
\caption{Long-term evolution of the planet HD\,11964\,c for an initial mutual
inclination of $30^\circ$ (Table\,\ref{T11964}). During the first million years,
the rotation rate is slowed down into its equilibrium position
(Eq.\,\ref{090520a}) and the obliquity is trapped in a small obliquity Cassini
state (Eq.\,\ref{090519ebis}). As the system evolves, 
tidal effects decrease the mutual inclination of the planetary orbits, while
the eccentricity increases
until around 2\,Gyr, time after which it is damped to zero.
The rotation rate closely follows the eccentricity, and the obliquity also
decreases, as the equilibrium in a Cassini state depends on the inclination of
the perturber. \llabel{F11964if}}
\end{figure*}

During the first million years (Figure\,\ref{F11964if}a), the rotation rate of
the inner planet and its obliquity are brought to
their equilibrium positions, where one believe that the spin of the planet can
be found today.
The rotation rate is slowed down until it reaches an equilibrium near 
$ \omega_\s / n_\b \approx f_2(e_\b) / f_1(e_\b) $, that is, $ P_\mathrm{rot,\s}
\approx 25 $\,d (Eq.\,\ref{110104b}), while
the obliquity, after an initial increase (since the initial rotation rate is
fast (Eq.\ref{090520d})), quickly drops until it is captured in the
small obliquity Cassini state $\de_\s \approx \sin I / \lambda_\s \approx 0.018^\circ$ (Eq.\ref{090519ebis}).
%If the initial obliquity is greater than $90^\circ$, capture in a large obliquity
%Cassini state may occur, but it is only temporary, since tidal torques disrupt
%this equilibrium as the rotation rate slows down \citep{Levrard_etal_2007} 
%(obs: standard simulation (m or x) with obl1=90.01)
Because the eccentricity and the inclination are oscillating, so does the
equilibrium rotation rate (Eq.\,\ref{090520a}) and the equilibrium obliquity 
(Eq.\,\ref{090519ebis}).

During the first two million years (Figure\,\ref{F11964if}a), the eccentricity and
the inclination of the inner orbit do not experience any substantial secular
modification (apart the periodical ones), since tidal effects are much less
efficient over the orbit than over the spin.
However, when we follow the system over Gyr time-scales 
one can observe those variations occurring (Figure\,\ref{F11964if}b).
The most striking effect is an initial increase in the eccentricity of the inner
planet, that is accompanied by a significant reduction in the mutual
inclination between the orbits of the two planets.
{\bfx
This behavior results from a combination of tidal effects, rotational flattening,
and the gravitational perturbations from the outer planet, which tend to
conserve the quantity $ h_1 = \sqrt{1-e_\b^2} \cos I $ (Eq.\,\ref{101221d}).
Tidal effects alone also decrease the eccentricity (Eq.\,\ref{090515c}),
so when its contribution becomes dominating the eccentricity is progressively
damped to zero.

The exchanges between eccentricity and inclination (Fig.\,\ref{nbodysims}) are
more significant for large values of the initial mutual inclination.
The maximum value for the eccentricity oscillations is then higher and therefore
the planet experiences stronger tidal effects.
As a result, the orbit of the inner planet evolves in a shorter time-scale.
In Figure\,\ref{F11964z} we plot the simultaneous evolution of the inclination
and eccentricity of the inner orbit for different initial mutual inclinations,
$I$.
For $ I < 30^\circ $ there is almost no reduction in the inclination, and the
eccentricity damping is very slow (Figure\,\ref{F11964z}a).
On the other hand, for $ I > 40^\circ $ we observe a reduction of more than
$20^\circ$ in the inclination, which is accompanied by an initial increase in
the eccentricity, followed by a rapid damping to zero
(Figure\,\ref{F11964z}c,d).
}

\begin{figure*}
\centering
\includegraphics[width=16cm]{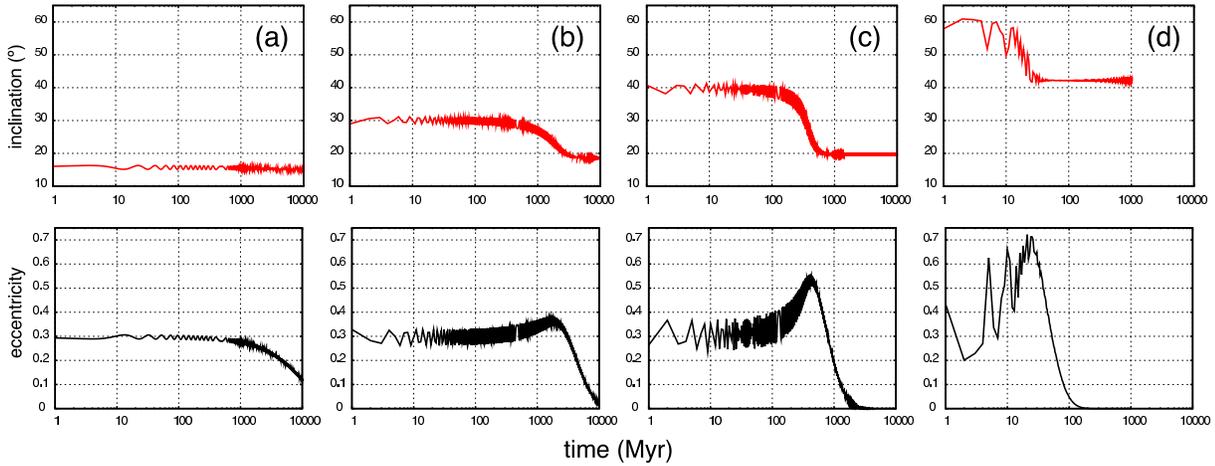}
\caption{Long-term evolution of the inclination (top) and eccentricity (bottom)
of the planet HD\,11964\,c for different values of the initial mutual
inclination: $15^\circ$ (a), $30^\circ$ (b), $40^\circ$ (c), and $60^\circ$ (d)
(Table\,\ref{T11964}). As the initial inclination increases, so does the
amplitude of the eccentricity oscillations, resulting that tidal effects are
enhanced, and the evolution time-scale shorter. \llabel{F11964z}}
\end{figure*}

\begin{figure*}
\centering
\includegraphics[width=15.2cm]{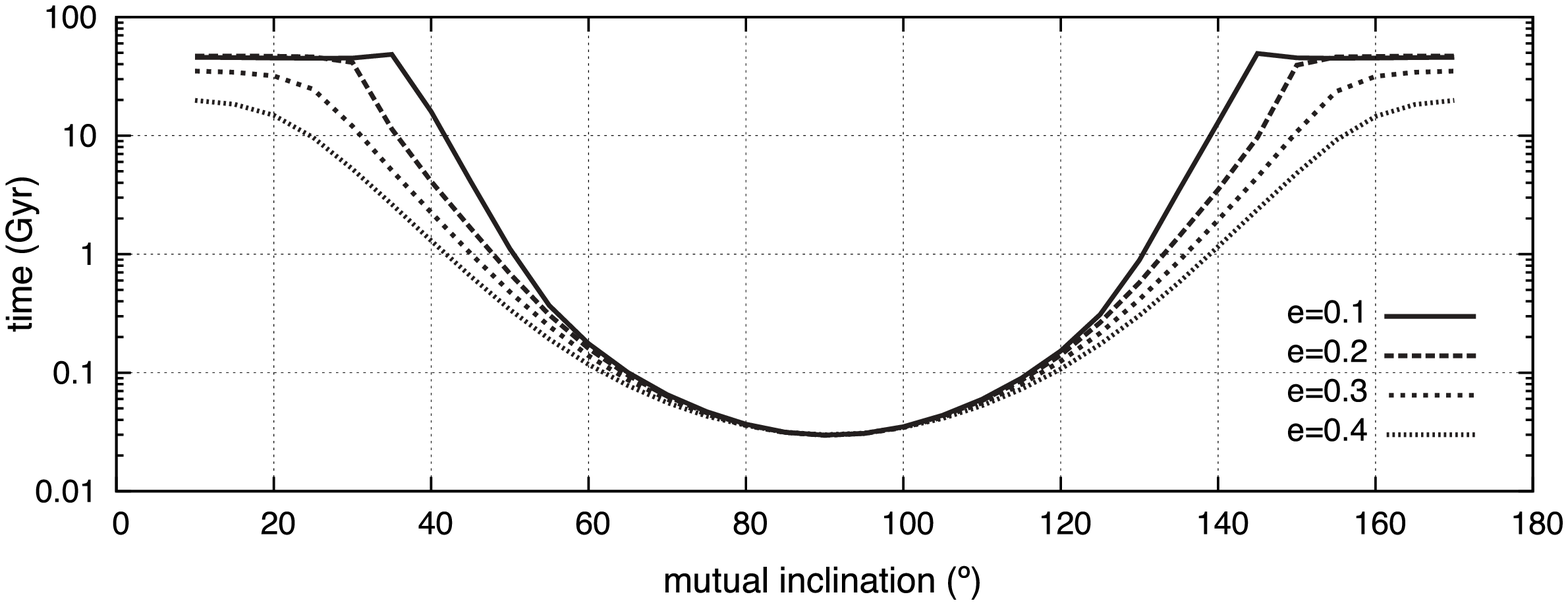}
\caption{Time needed to circularize the orbit of the planet HD\,11964\,c ($ e_\b
< 0.01 $) as a function of the initial mutual inclination and for different
values of the initial eccentricity ($ e_\b = 0.1$, 0.2, 0.3, and 0.4). 
The system is 10\,Gyr old \citep{Saffe_etal_2005}, so
we conclude that initial inclinations between $40^\circ$ and $140^\circ$ were not
possible, as the present eccentricity is still 0.3. \llabel{damping}}
\end{figure*}

For large values of the initial inclination,
the system evolves much faster to its final configuration, so the orbit
becomes circular in a shorter time-scale.
Since the present orbit of the inner planet still presents an eccentricity close
to 0.3, one may assume that the initial inclination could not have been excessively
large.
In order to estimate an upper limit for the initial inclination, we can run some
simulations and check which of those can evolve to the present configuration
within 10\,Gyr, the estimated age for the system. % \citep{Saffe_etal_2005}.

However, the choice of the initial orbit is not easy, since many different
initial configurations can bring the planet to the present orbit, 
depending on the tidal damping coefficients ($k_2$ and $\Delta t$) and also on
the initial eccentricity, inclination, and semi-major axis.
It is out of the scope of this study to explore exhaustively the initial
configurations of the HD\,11964 system, but rather to illustrate some interesting
evolutionary scenarios.
As we just have seen, the present eccentricity combined with a large inclination 
can later evolve to an identical eccentricity but with smaller inclination. 
The semi-major axis will be slightly smaller than today, but this parameter
essentially plays over the evolution time-scale. 
Thus, the present system can be a representation of the initial system and we
adopt it for simplicity.
In Figure\,\ref{damping} we used the initial conditions listed in
Table\,\ref{T11964}, but modified the initial inclinations and eccentricities.
We observe that inclinations in the interval $[40^\circ, 140^\circ]$ circularize
the orbit ($ e_\b < 0.01 $) in less than 10\,Gyr, so they can be discarded.

%\vskip1.truecm
\section{Conclusion}

Many multi-planet systems have been reported in hierarchical configurations.
For the most part, their mutual inclinations are unknown, but the fact that they
exhibit significant values in the eccentricities led to think that the
inclinations can also be large.
In addition, very often the innermost planet in these systems is very
close to the star and undergoes tidal evolution.
Here we have studied this particular sub-group of multi-planet systems.
Using a very general and simplified averaged vectorial formalism, we have shown that
inclined hierarchical planetary systems undergoing tidal dissipation can evolve in many
different and sometimes unexpected ways.

We re-analyzed the case of HD\,80606, a system where the planet migrated inwards
by a combined tidal-Kozai mechanism, confirming previous results.
We looked at the behavior of a planet (or disc) around the binary
stars HD\,98800\,B and showed that initial prograde orbits may become retrograde
and vice-versa, only because of tidal migration within the binary stars.
Finally, we studied the regular 2-planet HD\,11964 system and showed that tidal
dissipation combined with gravitational perturbations may lead to a decrease in
the mutual inclination, and a fast circularization of the inner orbit.

We have chosen the above three examples, as they are representative of the
diversity of behaviors among inclined hierarchical systems.
Many other systems are awaiting to be studied.
The fact that we use average equations for both tidal and gravitational effects,
makes our method suitable to be applied in long-term studies.
It allows to run many simulations for different initial conditions in order to
explore the entire phase space and evolutionary scenarios.
In particular, it can be very useful to put constraints on the inclinations and
dissipation ratios of hierarchical planetary systems.
{\bfx
Our study can also be extended 
to systems of binary stars, and to planet-satellite systems.
}

\begin{acknowledgements}
We acknowledge support from PNP-CNRS, France, and from Funda\c{c}\~ao
para a Ci\^encia e a Tecnologia, Portugal (grant PTDC/CTE-AST/098528/2008).
\end{acknowledgements}

\appendix

\section{Averaged quantities}

\llabel{apenA}

For completeness, we gather here the average formulae that are used in
the computation of secular equations. Let $F(\vv{r},\dot \vv{r})$ be
a function of a position vector $\vv{r}$ and velocity $\dot \vv{r}$. Its
averaged expression over the mean anomaly $(M)$ is given by
\be
\moy{F}_{M} = \frac{1}{2\pi}\int_0^{2\pi} F(\vv{r}, \dot \vv{r})\, \d M\ .
\ee
Depending on the case, this integral is computing using the eccentric
anomaly $(E)$, or the true anomaly $(v)$ as an intermediate variable. The
basic formulae are
\begin{eqnarray}
&\d M = \Frac{r}{a}\d E = \Frac{r^2}{a^2\sqrt{1-e^2}}\d v\ , \crm
&\vv{r} = a(\cos E-e)\, \hat \vv{e} + a\sqrt{1-e^2}(\sin E)\, \hat \vv{k}
\times \hat \vv{e}\ , \crm
&\vv{r} = r\cos v\, \hat \vv{e} + r\sin v\, \hat \vv{k} \times \hat
\vv{e}\ , \crm
&\dot \vv{r} = \Frac{na}{\sqrt{1-e^2}}\, \hat \vv{k} \times 
(\hat \vv{r} + \vv{e})\ , \crm
&r = a (1-e\cos E) = \Frac{a(1-e^2)}{1+e\cos v}\ ,
\end{eqnarray}
where $\hat \vv{k}$ is the unit vector of the orbital angular momentum,
and $\vv{e}$ the Laplace-Runge-Lenz vector (Eq.\,\ref{100119a}). 
We have then
\be
\moy{\Frac{1}{r^3}} = \Frac{1}{a^3(1-e^2)^{3/2}}\ ,
\ee
and
\be
\moy{\frac{\vv{r} \trans{\vv{r}}}{r^5}} = 
\Frac{1}{2a^3(1-e^2)^{3/2}} \left(1-\hat \vv{k} \trans{\hat \vv{k}} \right) \ ,
\ee
which leads to
\be
\moy{\Frac{1}{r^3}P_2(\hat \vv{r} \cdot \hat \vv{u})}
= -\Frac{1}{2a^3(1-e^2)^{3/2}} P_2(\hat \vv{k} \cdot \hat \vv{u})\ ,
\ee
for any unit vector $\hat \vv{u}$. In the same way,
\be
\moy{r^2} = a^2\left(1+\frac{3}{2}e^2\right)\ ,
\ee
and
\be
\moy{\vv{r} \trans{\vv{r}}} = a^2\Frac{1-e^2}{2}
\left(1-\hat \vv{k} \trans{\hat \vv{k}}\right) + \frac{5}{2} a^2 \vv{e}
\trans{\vv{e}}\ ,
\ee
give
\be
\moy{r^2 P_2(\hat \vv{r} \cdot \hat \vv{u})} = 
-\Frac{a^2}{2}\Big((1-e^2)P_2(\hat \vv{k} \cdot \hat \vv{u}) - 
5 e^2 P_2(\hat \vv{e} \cdot \hat \vv{u})\Big)\ .
\ee
The other useful formulae are
\be
\moy{\Frac{1}{r^6}} = \frac{1}{a^6} f_1(e)\ ,
\ee
\be
\moy{\Frac{1}{r^8}} = \frac{1}{a^8\sqrt{1-e^2}} f_2(e)\ ,
\ee
\be
\moy{\Frac{\vv{r} \trans{\vv{r}}}{r^8}} = 
\Frac{\sqrt{1-e^2}}{2a^6} f_4(e) \left(1-\hat \vv{k} \trans{\hat \vv{k}}\right)
+\frac{6+e^2}{4a^6(1-e^2)^{9/2}} \vv{e} \trans{\vv{e}}\ ,
\ee
\be
\moy{\Frac{\vv{r}}{r^8}} = \frac{5}{2}\Frac{1}{a^7\sqrt{1-e^2}}\, f_4(e)
\vv{e}\ ,
\ee
\be
\moy{\Frac{\vv{r}}{r^{10}}} = \frac{7}{2}\Frac{1}{a^9(1-e^2)}\, f_5(e)
\vv{e}\ ,
\ee
\be
\moy{\Frac{(\vv{r} \cdot \dot \vv{r}) \vv{r}}{r^{10}}} = \Frac{n}{2a^7
\sqrt{1-e^2}} f_5(e)\, \hat \vv{k} \times \vv{e}\ ,
\ee
where the $f_\ij (e)$ functions are given by expressions (\ref{090514n}) to
(\ref{090515e}).

Finally, for the average over the \longitude of the periapsis ($\varpi$), 
%that we used to obtain expression (\ref{090514k}), 
we can proceed in an identical manner:
\be
\moy{\vv{e} \trans{\vv{e}}}_\varpi = \frac{1}{2 \pi} \int_0^{2 \pi} \vv{e}
\trans{\vv{e}} \, d \varpi = \frac{e^2}{2} \left( 1 - \vv{k} \trans{\vv{k}}
\right) \ ,
\ee
which gives
\be
\moy{ \left( \vv{e} \cdot \hat \vv{u} \right) \vv{e}}_\varpi = \frac{e^2}{2}
\Big( \hat \vv{u} - ( \vv{k} \cdot \hat \vv{u} ) \vv{k} \Big) \ .
\ee

\bibliographystyle{spbasic}      % basic style, author-year citations
\bibliography{correia}   % name your BibTeX data base

\end{document}